\documentclass[twocolumn]{aastex631}
\usepackage{natbib}
\usepackage{rotfloat}
\hypersetup{linkcolor=blue,citecolor=blue,filecolor=cyan,urlcolor=magenta}

\newcommand{\Msun}{\mathrm{M}_{\odot}}

\newcommand{\Ha}{H$\alpha$}
\newcommand{\Hb}{H$\beta$}

\newcommand{\OIII}{[{O\,}{\scriptsize {III}}]}
\newcommand{\NII}{[{N\,}{\scriptsize {II}}]}

\newcommand{\OII}{[{O\,}{\scriptsize {II}}]}

\newcommand{\SII}{[{S\,}{\scriptsize {II}}]}
\newcommand{\Lya}{Ly$\alpha$}

\accepted{by ApJ}

\shorttitle{THE PROPERTIES OF \OIII\ ELGs AT $z=3.25$}
\shortauthors{Wen et al.}
\graphicspath{{./}}
\begin{document}

\defcitealias{An2014}{A14}

\title{The Physical Properties of Star-Forming Galaxies with Strong \OIII\ Lines at $z=3.25$}

\author{ Run Wen }
\affiliation{Purple Mountain Observatory, Chinese Academy of Sciences, 10 Yuanhua Road, Qixia District, Nanjing 210023, China}
\affiliation{School of Astronomy and Space Sciences, University of Science and Technology of China, Hefei 230026, China}

\author{ FangXia An }
\affiliation{Department of Physics and Astronomy, University of the Western Cape, and the Inter-University Institute for Data Intensive Astronomy, Robert Sobukwe Road, 7535 Bellville, Cape Town, South Africa}

\author{ Xian~Zhong Zheng }
\affiliation{Purple Mountain Observatory, Chinese Academy of Sciences, 10 Yuanhua Road, Qixia District, Nanjing 210023, China}
\affiliation{School of Astronomy and Space Sciences, University of Science and Technology of China, Hefei 230026, China}
\email{xzzheng@pmo.ac.cn}

\author{ Dong~Dong Shi }
\affiliation{Purple Mountain Observatory, Chinese Academy of Sciences, 10 Yuanhua Road, Qixia District, Nanjing 210023, China}

\author{ Jianbo Qin }
\affiliation{Purple Mountain Observatory, Chinese Academy of Sciences, 10 Yuanhua Road, Qixia District, Nanjing 210023, China}

\author{ Valentino Gonzalez }
\affiliation{Chinese Academy of Sciences South America Center for Astronomy, China-Chile Joint Center for Astronomy, Camino del Observatorio 1515, Las Condes, Chile}
\affiliation{Centro de Astrofísica y Tecnologías Afines (CATA), Camino del Observatorio 1515, Las Condes, Santiago, Chile}

\author{ Fuyan Bian }
\affiliation{European South Observatory, Alonso de Cordova 3107, Casilla 19001, Vitacura, Santiago 19, Chile}

\author{ Haiguang Xu }
\affiliation{School of Physics and Astronomy, Shanghai Jiao Tong University, 800 Dongchuan Road, Shanghai 200240, China}

\author{ Zhizheng Pan }
\affiliation{Purple Mountain Observatory, Chinese Academy of Sciences, 10 Yuanhua Road, Qixia District, Nanjing 210023, China}

\author{ Qing-Hua Tan }
\affiliation{Purple Mountain Observatory, Chinese Academy of Sciences, 10 Yuanhua Road, Qixia District, Nanjing 210023, China}

\author{ Wenhao Liu }
\affiliation{Purple Mountain Observatory, Chinese Academy of Sciences, 10 Yuanhua Road, Qixia District, Nanjing 210023, China}

\author{ Min Fang }
\affiliation{Purple Mountain Observatory, Chinese Academy of Sciences, 10 Yuanhua Road, Qixia District, Nanjing 210023, China}

\author{ Jian Ren }
\affiliation{Purple Mountain Observatory, Chinese Academy of Sciences, 10 Yuanhua Road, Qixia District, Nanjing 210023, China}
\affiliation{School of Astronomy and Space Sciences, University of Science and Technology of China, Hefei 230026, China}

\author{ Yu~Heng Zhang }
\affiliation{Purple Mountain Observatory, Chinese Academy of Sciences, 10 Yuanhua Road, Qixia District, Nanjing 210023, China}
\affiliation{School of Astronomy and Space Sciences, University of Science and Technology of China, Hefei 230026, China}

\author{ Man Qiao }
\affiliation{Purple Mountain Observatory, Chinese Academy of Sciences, 10 Yuanhua Road, Qixia District, Nanjing 210023, China}
\affiliation{School of Astronomy and Space Sciences, University of Science and Technology of China, Hefei 230026, China}

\author{ Shuang Liu }
\affiliation{Purple Mountain Observatory, Chinese Academy of Sciences, 10 Yuanhua Road, Qixia District, Nanjing 210023, China}
\affiliation{School of Astronomy and Space Sciences, University of Science and Technology of China, Hefei 230026, China}



\begin{abstract}

We present an analysis of physical properties of 34 \OIII\ emission-line galaxies (ELGs) at $z=3.254\pm0.029$ in the Extended Chandra Deep Field South (ECDFS).  These ELGs are selected from deep narrow H$_2\mathrm{S}(1)$ and broad $K_\mathrm{s}$ imaging of 383\,arcmin$^2$ obtained with CFHT/WIRCam.  We construct spectral energy distributions (SEDs) from $U$ to $K_\mathrm{s}$ to derive the physical properties of ELGs.  These \OIII\ ELGs are identified as starburst galaxies with strong \OIII\ lines of $L_\mathrm{\OIII}\sim10^{42.6}-10^{44.2}$\,erg\,s$^{-1}$, and have stellar masses of $M_{\ast}\sim$\,10$^{9.0}-10^{10.6}$\,M$_\odot$ and star formation rates of $\sim$\,10--210\,M$_\odot$\,yr$^{-1}$.  Our results show that 24\% of our sample galaxies are dusty with $A_V>1$\,mag and EW(\OIII)$_\mathrm{rest}\sim$\,70--500\,\AA, which are often missed in optically selected \OIII\ ELG samples.  Their rest-frame UV and optical morphologies from \textit{HST}/ACS and \textit{HST}/WFC3 deep imaging reveal that these \OIII\ ELGs are mostly multiple-component systems (likely mergers) or compact.  And 20\% of them are nearly invisible in the rest-frame UV owing to heavy dust attenuation.   Interestingly, we find that our samples reside in an overdensity consisting of two components: one southeast (SE) with an overdensity factor of $\delta_{\rm gal}\sim$41 over a volume of 13$^3$\,cMpc$^3$ and the other northwest (NW) with $\delta_{\rm gal}\sim$38 over a volume of 10$^3$\,cMpc$^3$.  The two overdense substructures are expected to be virialized at $z=0$ with a total mass of $\sim 1.1\times10^{15}\,$M$_\odot$ and $\sim 4.8\times10^{14}\,$M$_\odot$, and probably merge into a Coma-like galaxy cluster. 
\end{abstract}

\section{Introduction} \label{sec:intro}

The past two decades have witnessed a wealth of progress in mapping galaxy formation and evolution.  The current generation of multiwavelength deep surveys have revealed the detailed properties of galaxy populations out to $z\sim2$--3, where the cosmic star formation rate density (CSFRD) reaches its peak \citep{Hopkins2006,Sobral2013,Madau2014,Khostovan2015}.  At $z>2$--3 about one-quarter of the present-day stars were formed in the progenitors of present-day massive galaxies \citep{Madau2014}, preferentially in the overdense environments  \citep{Thomas2005,Chiang2017}.  Characterizing the properties of galaxies at $z>3$ is thus essential to understanding the early formation of massive galaxies and large-scale structures, as well as how the star formation activities are activated to reach the peak of CSFRD \citep{Suzuki2015,Onodera2016,Onodera2020}.

The emission lines in the rest-frame optical spectra of galaxies (e.g., \OII$\lambda\lambda$3727, 3729, \Hb, \OIII$\lambda\lambda$4959, 5007, \NII$\lambda\lambda$6549, 6585, \Ha\ and \SII$\lambda\lambda$6718, 6732) are mostly used for physical and chemical diagnostics \citep[see][for a review]{Kewley2019}.  Moreover, studies of emission-line galaxies (ELGs) at $z>3$ provide insights into understanding the cosmic reionization \citep{deBarros2016}.  The universe is fully ionized by $z\sim6$ \citep[e.g.,][]{Fan2006,deBarros2014}.  Star-forming galaxies (SFGs) at $z>6$ are thought to be the main contributors to the ionizing field in the era of reionization \citep[e.g., ][]{Nakajima2014,Robertson2015}.  Owing to the opaque intergalactic medium \citep[e.g., ][]{Worseck2014}, the nature of ionizing sources in the reionization era is still not well understood.

These ionizing sources usually have prominent \OIII+\Hb\ emission \citep{deBarros2019,Endsley2021}.  The strong \OIII\ emission lines may reveal the extreme conditions of the interstellar medium in a galaxy, and likely are associated with low metallicity and high ionizing parameters \citep{McLinden2011,Nakajima2014,Onodera2020,Tang2021b}.  The extreme \OIII\ ELGs are often seen as analogs of galaxies in the reionization era \citep{Tang2019,Du2020,Tang2021a,Tang2022}. And galaxies with large \OIII\ equivalent widths (EWs; from 200\,\AA\ to 800\,\AA) are widely used to address the \Lya\ continuum escape fraction in the high-$z$ universe \citep{Fletcher2019,Barrow2020,Katz2020,Nakajima2020}.  Their analogs at low $z$ refer to the so-called “Green Pea” galaxies \citep{Cardamone2009}, showing strong \OIII\ emission with extremely high \OIII$/$\OII\ ratio \citep{Jaskot2013,Yang2017,Yuma2019,Lumbreras-Calle2021,Liu2022}.  The \OIII\ lines redshift into the near-infrared (NIR) and mid-infrared (MIR) bands for $z>3$ objects.  Deep IR photometric and spectroscopic observations are thus crucial to identifying and studying SFGs at $z>3$ \citep[e.g.,][]{Bunker1995,Geach2008,Nakajima2013,Sobral2013,Khostovan2016}.

However, the NIR observations of high-$z$ galaxies can be carried out only in the $J$, $H$ and $K_\mathrm{s}$ bands on the ground owing to the atmospheric transmission.  And the high sky background leads such observations to be very time-consuming and available only for a limited sky area.  MOSFIRE on board the Keck telescope is an efficient instrument in taking NIR spectroscopy of high-$z$ galaxies \citep{McLean2012}.  The NIR spectroscopic surveys with MOSFIRE, e.g., the Keck Baryonic Structure Survey \citep[KBSS-MOSFIRE;][]{Steidel2014} and the MOSFIRE Deep Evolution Field survey \citep[MOSDEF;][]{Kriek2015,Shapley2015,Reddy2018}, obtained rest-frame optical spectra of thousands of galaxies mostly at $z\sim1.4$--3 based on the $H$-band selection.  In contrast, the NIR observations with space-borne facilities are free from the atmospheric emission, but constrained by the thermal emission from the facilities.  The IR grism surveys using WFC3 on board the \textit{Hubble Space Telescope} (\textit{HST}), e.g., the WFC3 Infrared Spectroscopic Parallel Survey \citep[WISP;][]{Atek2011}, the MAMMOTH-Grism HST slitless spectroscopic survey \citep{Wang2021} and the 3D-HST survey \citep{Brammer2012,Momcheva2016}, provided low-resolution rest-frame optical spectra for a large number of galaxies out to $z\sim2.5$.  These NIR surveys have built a more comprehensive view of the physical properties of galaxies at $1<z<3$.  The \textit{James Webb Space Telescope} (JWST) will offer unprecedented sensitivities in the NIR and MIR to conduct imaging and spectroscopy of high-$z$ galaxies and revolutionize our understanding of galaxy formation and evolution since the era of reionization.

Deep imaging through narrow- and broadband $K_\mathrm{s}$ enables us to detect the emission lines \OIII$\lambda\lambda$4959,5007 in galaxies over $3<z<3.7$, and even determine the \OIII\ luminosity functions \citep{Reddy2008, Kashikawa2011, Khostovan2015, Sobral2015, Gong2017, Khostovan2020}.  Such observations are often used to identify \Ha\ and other emission lines at lower redshifts \citep{Khostovan2015, Khostovan2016}.  It has been verified that the approach with NIR narrowband imaging is effective in probing ELGs within a narrow redshift range of $\delta z/(1+z)=$1\%--2\% over a large sky coverage \citep{Sobral2013}.  On the other hand, the presence of strong \OIII\ emission lines may cause an excess of the observed $K_\mathrm{s}$ flux relative to the continuum flux derived from broadband spectral energy distributions (SEDs) and be used to identify \OIII\ ELGs at $3<z<3.7$ \citep{Onodera2020}.  Pilot studies of \OIII\ ELGs with spectroscopic observations have been contributed to addressing the kinematic and structural evolution of \OIII\ SFGs \citep{Steidel2010,McLinden2013,Schenker2013,Gillman2019,Price2020,Tran2020,Yates2020}, as well as metal enrichments \citep[e.g.,][]{Kewley2008,Mannucci2010,Sommariva2012,Nakajima2014,Troncoso2014,Nakajima2016}.

In the Extended Chandra Deep Field South (ECDFS), a deep narrowband imaging survey has been carried out with CFHT/WIRCam, detecting a sample of 34 \OIII\ ELGs at $z\sim 3.25$ \citep[][hereafter A14]{An2014}.  Here we conduct a detailed analysis of the physical properties of these \OIII\ ELGs using the publicly available multiwavelength data.  This paper is organized as follows: In Section~\ref{sec:observation}, we briefly introduce our narrowband imaging observations and multiwavelength data used in our analysis.  Section~\ref{sec:sedfitting} displays the input parameters of SED fitting and gives the results.  We present the physical properties of our \OIII\ samples in Section~\ref{sec:results}, and the overdensity in ECDFS traced by \OIII\ is shown in Section~\ref{sec:overdensity}.  We discuss about our results in Section~\ref{sec:discussion}.  In Section~\ref{sec:conclusion} we give a summary of this work.   Throughout this paper we adopt cosmological parameters of $\Omega_\mathrm{M}=0.3$, $\Omega_\mathrm{\Lambda}=0.7$, and H$_0 = 70$\,km\,s$^{-1}$\,Mpc$^{-1}$.  Unless otherwise stated, all magnitudes are given in the AB magnitude system \citep{Oke1974}, and a Chabrier initial mass function \citep[IMF;][]{Chabrier2003} is assumed.

\begin{figure}
\centering
\includegraphics[width=0.45\textwidth]{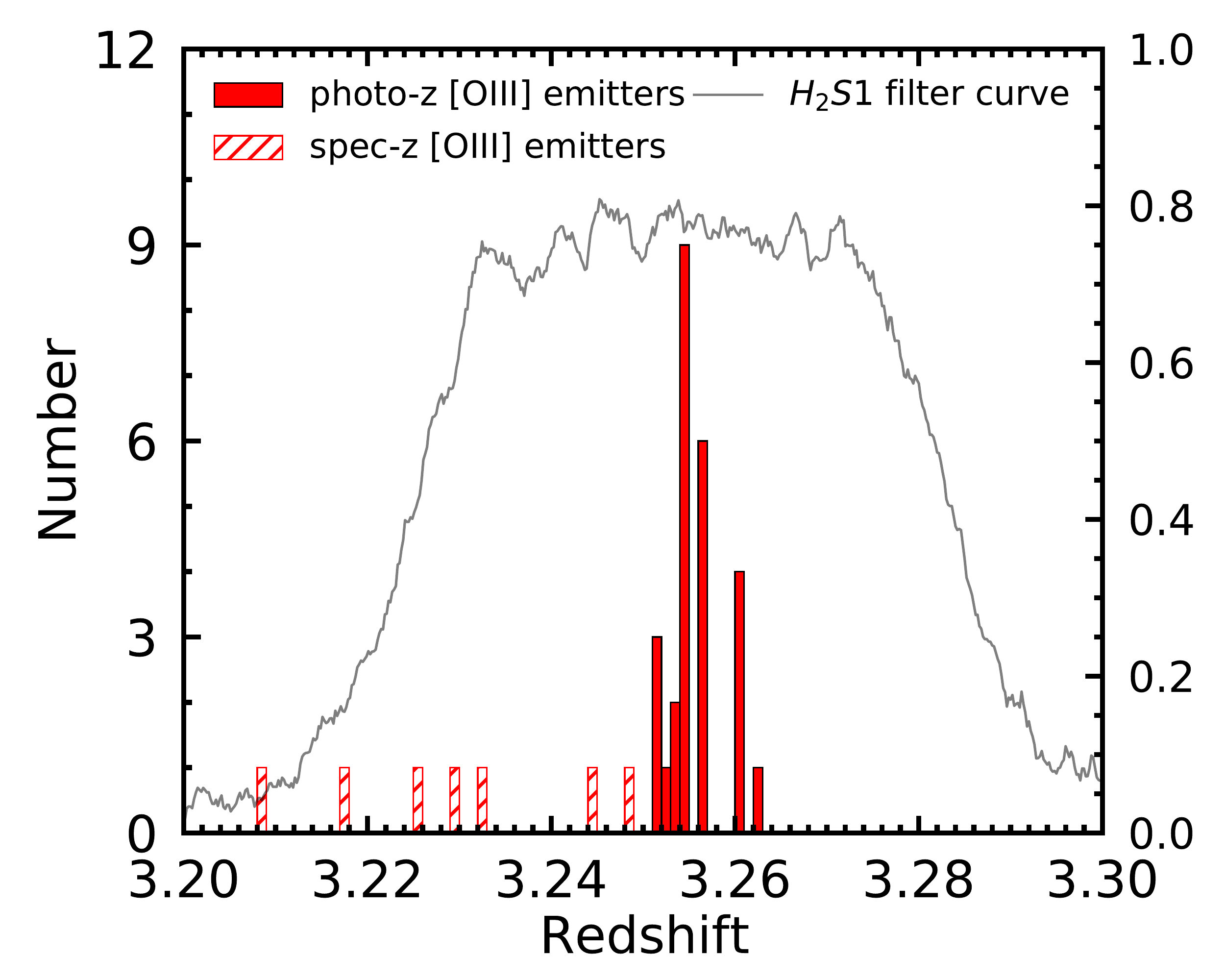}
\caption{Redshift distribution of 34 \OIII\ ELGs. A bin width of 0.001 is adopted. 
	The filled histogram presents the distribution of \OIII\ ELGs with photo-$z$ 
	while the hatched histogram refers to spec-$z$.
	The gray curve is the width range of H$_2$S(1) filter for the \OIII$\lambda$5007 emission line at $z\sim3.25$.}
\label{f:zdistribution.pdf}
\end{figure}

\section{Sample and Data}\label{sec:observation}

\citetalias{An2014} presented the observations and detections of 140 emission-line objects with the H$_2\mathrm{S}(1)$ narrowband and $K_\mathrm{s}$ broadband imaging of ECDFS.  Of the 140 objects, 34 are recognized as \OIII\ ELGs at $z\sim3.25$.  Here we briefly describe the NIR observations, data, and selection for the sample of 34 \OIII\ ELGs.  More details can be found in \citetalias{An2014}.

\subsection{Sample Selection}\label{sec:data}

A deep imaging survey of ECDFS (centered at $\alpha$=03:28:45, $\delta$=$-$27:48:00) was conducted through the narrowband filter H$_2\mathrm{S}(1)$ ($\lambda_\mathrm{c}=2.130\,\micron$, $\Delta\lambda=0.0293\,\micron$) with the instrument WIRCam on board the Canada-France-Hawaii Telescope \citep[CFHT; ][]{Puget2004}.  WIRCam consists of four 2048 $\times$ 2048 HAWAII2-RG detectors, providing a sky coverage of $20\arcmin\times20\arcmin$ with a pixel scale of $0\farcs3$  pixel$^{-1}$.  The deep $K_\mathrm{s}$-band ($\lambda_\mathrm{c}=2.146\,\micron$, $\Delta\lambda=0.325\,\micron$) imaging data were also obtained with CFHT/WIRCam and adopted in our analysis from \citet{Hsieh2012}.  The final science images reach a 5$\sigma$ depth of of H$_2\mathrm{S}(1)=22.8$\,mag and $K_\mathrm{s}=24.8$\,mag for point sources.

The H$_2\mathrm{S}(1)$ and $K_\mathrm{s}$ imaging data of ECDFS are used to identify ELGs with the  $K_\mathrm{s} - \mathrm{H}_2\mathrm{S}(1)$ color excess \citep{Bunker1995} following 
\begin{equation}  \label{equ:selection}
	 K_\mathrm{s}-H_2\mathrm{S}(1) > -2.5\,\log(1 - \Sigma\,\sqrt{\sigma_{K_\mathrm{s}}^2 + \sigma^2_{\mathrm{H_2\mathrm{S}(1)}}}/f_{\mathrm{H_2\mathrm{S}(1)}} \,), 
\end{equation}
where $\Sigma$ is the significant factor and $\sigma_{\rm H_2\mathrm{S}(1)}$ and $\sigma_{K_{\rm s}}$ are background noises in the two bands.  The $\Sigma$ is introduced to quantify the significance of a narrowband excess relative to the combined 1$\sigma$ photometric error from both the narrow- and broadband. Here $f_\mathrm{H_2\mathrm{S}(1)}$ refers to H$_2\mathrm{S}(1)$ flux as $f_\mathrm{H_2\mathrm{S}(1)}=0.3631\times10^{0.4\,(25 - H_2\mathrm{S}(1))}$.  The noises and fluxes are given in units of $\mu$Jy.  Using the color excess criteria, in total 8720 sources were detected with S/N$>5$ in both H$_2\mathrm{S}(1)$ and $K_\mathrm{s}$.  Their fluxes were measured from the corresponding images using the software tool SExtractor \citep{Bertin1996}.  With the selection criteria of $\Sigma=3$ and EW$>50$\AA, in total 140 objects were securely selected as emission-line candidates.

\subsection{Public Data} \label{sec:sample}

We utilize optical $U$-, $B$-, $V$-, $R$- and $I$-band photometric catalog and imaging data from the Multiwavelength Survey by Yale-Chile \citep[MUSYC;][]{Gawiser2006,Cardamone2010}; \textit{HST}/ACS F606W ($V_\mathrm{606}$) and F850LP ($z_\mathrm{850}$) imaging from the GEMS survey \citep{Rix2004,Caldwell2008}; \textit{HST}/WFC3 F125W ($J_\mathrm{125}$) and F160W ($H_\mathrm{160}$) imaging from the Cosmic Assembly Near-infrared Deep Extragalactic Legacy Survey \citep[CANDELS;][]{Grogin2011,Koekemoer2011}; and CFHT/WIRCam $J$ and $K_\mathrm{s}$ imaging data \citep{Hsieh2012}, in conjunction with our H$_2\mathrm{S}(1)$ imaging data.  Fluxes in these 12 bands are obtained for the sample of 140 emission-line candidates.  Note that 16 targets are optically too faint to be included in the MUSYC public catalog, and 72 sources out of the 140 candidates have $J_\mathrm{125}$ and $H_\mathrm{160}$ data since the CANDELS survey only covers a part of ECDFS.

By fitting their SEDs composed of 12-band data points with the software tool EAZY \citep[Easy and Accurate Redshifts from Yale; ][]{Brammer2008}, photometric redshifts (photo-$z$) were derived for the 140 emission-line candidates (the modeling will be introduced in Section~\ref{sec:sedfitting}).  Of them, 34 ELGs with $2.8<z_\mathrm{phot}<3.3$ are identified as \OIII\ emitters at $z\sim3.25$.  With the public catalogs available in the literature, we also identify that 8 of these 34 ELGs have spectroscopic redshifts (spec-$z$).  Except for one source that has a spec-$z$ at $3.083$, the spec-$z$ of the remaining seven sources is in range of 3.208--3.248, confirming that these ELGs are located at $z\sim3.25$. 

\begin{longrotatetable}
%\movetabledown=0in
\begin{deluxetable*}{ccccccccccccccccc}
\tablecaption{Multiband Photometry of 34 \OIII\ Emitters of Our Samples \label{tab:photometry}}
\tabcolsep=3.5pt
\tablewidth{640pt}
\tabletypesize{\scriptsize}
\tablehead{
	\colhead{ID} & \colhead{R.A.(J2000)} & \colhead{Decl.(J2000)} & \colhead{$f_U$} & \colhead{$f_B$} & 
\colhead{$f_V$} & \colhead{$f_R$} & \colhead{$f_I$} & \colhead{$f_{v606}$} &
\colhead{$f_{z850}$} & \colhead{$f_{125}$} & \colhead{$f_{160}$} & \colhead{$f_J$} &
\colhead{$f_{H_2S1}$} & \colhead{$K_s$} 
%& \colhead{$f_{3.6}$} & 
%\colhead{$f_{4.5}$} & 
%\\
%\colhead{} & \colhead{J2000.0} & \colhead{J2000.0} & \multicolumn{12}{c}{$\mu$ Jy}
}
\startdata
1 &  53.176239 & $-$27.978130 &  $-$0.15$\pm$0.02 &   0.20$\pm$0.01 &   0.38$\pm$0.02 &   0.53$\pm$0.11 &   0.54$\pm$0.02 &   0.45$\pm$0.01 &   0.70$\pm$0.05 & \nodata & \nodata &  0.69$\pm$0.07 &  2.60$\pm$0.41 &  1.23$\pm$0.11 \\
%&  1.25$\pm$0.08 &  0.77$\pm$0.10 \\
2 &  53.159058 & $-$27.972429 &  $-$0.13$\pm$0.02 &   0.18$\pm$0.01 &   0.32$\pm$0.02 &   0.44$\pm$0.10 &   0.49$\pm$0.02 &   0.40$\pm$0.01 &   0.62$\pm$0.04 & \nodata & \nodata &  0.80$\pm$0.06 &  4.14$\pm$0.33 &  1.69$\pm$0.09 \\
%&  2.25$\pm$0.10 &  2.98$\pm$0.13 \\
3 &  53.178284 & $-$27.969770 &  $-$0.17$\pm$0.02 &   0.39$\pm$0.02 &   0.65$\pm$0.02 &   0.98$\pm$0.12 &   0.84$\pm$0.02 &   0.74$\pm$0.02 &   0.93$\pm$0.05 & \nodata & \nodata &  1.00$\pm$0.10 &  6.93$\pm$0.47 &  2.06$\pm$0.14 \\
%&  2.77$\pm$0.09 &  2.54$\pm$0.10 \\
4 &  53.239136 & $-$27.951475 &  $-$0.20$\pm$0.03 &   0.15$\pm$0.02 &   0.30$\pm$0.02 &   0.47$\pm$0.14 &   0.46$\pm$0.02 &   0.38$\pm$0.02 &   0.46$\pm$0.06 & \nodata & \nodata &  0.49$\pm$0.08 &  3.09$\pm$0.46 &  1.20$\pm$0.12 \\
%&  1.22$\pm$0.07 &  1.18$\pm$0.09 \\
5 &  53.158218 & $-$27.950077 &  $-$0.10$\pm$0.01 &   0.36$\pm$0.01 &   0.65$\pm$0.01 &   1.30$\pm$0.08 &   0.99$\pm$0.01 & \nodata &   0.59$\pm$0.03 & \nodata & \nodata &  1.39$\pm$0.07 &  5.81$\pm$0.28 &  2.97$\pm$0.11 \\
%&  4.91$\pm$0.24 &  4.88$\pm$0.21 \\
6 &  52.965435 & $-$27.948175 &  $-$0.14$\pm$0.02 &   0.17$\pm$0.01 &   0.31$\pm$0.02 &   0.54$\pm$0.11 &   0.40$\pm$0.02 &   0.36$\pm$0.01 &   0.42$\pm$0.04 & \nodata & \nodata &  0.61$\pm$0.06 &  5.61$\pm$0.27 &  1.32$\pm$0.09 \\
%&  1.24$\pm$0.06 &  1.06$\pm$0.10 \\
7 &  53.173649 & $-$27.943214 &  $-$0.10$\pm$0.01 &  $-$0.06$\pm$0.01 &   0.10$\pm$0.01 &  $-$0.41$\pm$0.07 &   0.17$\pm$0.01 & \nodata &   0.29$\pm$0.03 & \nodata & \nodata &  0.20$\pm$0.05 &  2.65$\pm$0.22 &  0.70$\pm$0.07 \\
%&  2.65$\pm$0.14 &  1.32$\pm$0.26 \\
8 &  53.162315 & $-$27.942719 &  $-$0.10$\pm$0.01 &   0.17$\pm$0.01 &   0.32$\pm$0.01 &   0.65$\pm$0.08 &   0.45$\pm$0.01 & \nodata &   0.59$\pm$0.03 & \nodata & \nodata &  1.12$\pm$0.05 &  4.57$\pm$0.24 &  1.79$\pm$0.08 \\
%&  0.66$\pm$0.13 &  1.53$\pm$0.12 \\
9 &  53.171169 & $-$27.932827 &  $-$0.17$\pm$0.03 &   0.11$\pm$0.02 &   0.27$\pm$0.02 &  $-$0.72$\pm$0.13 &   0.34$\pm$0.02 &   0.30$\pm$0.00 &   0.67$\pm$0.01 &   0.48$\pm$0.02 &   0.79$\pm$0.02 &  0.38$\pm$0.06 &  2.97$\pm$0.37 &  0.97$\pm$0.09 \\
%&  1.48$\pm$0.07 &  1.43$\pm$0.09 \\
10 &  53.170116 & $-$27.929638 & \nodata & \nodata & \nodata & \nodata & \nodata &   0.17$\pm$0.01 &   0.36$\pm$0.02 &   1.07$\pm$0.02 &   2.24$\pm$0.02 &  0.95$\pm$0.05 & 26.41$\pm$0.28 &  8.25$\pm$0.08 \\
%& 13.94$\pm$0.07 & 21.58$\pm$0.09 \\
11 &  53.161907 & $-$27.920698 &  $-$0.12$\pm$0.02 &   0.08$\pm$0.01 &   0.13$\pm$0.02 &  $-$0.50$\pm$0.09 &   0.18$\pm$0.01 &   0.15$\pm$0.01 &   0.19$\pm$0.02 &   0.16$\pm$0.01 &   0.20$\pm$0.02 &  0.21$\pm$0.05 &  2.19$\pm$0.26 &  0.62$\pm$0.07 \\
%&  0.33$\pm$0.05 &  0.23$\pm$0.07 \\
12 &  53.167480 & $-$27.913368 &  $-$0.12$\pm$0.02 &  $-$0.07$\pm$0.01 &  $-$0.09$\pm$0.02 &  $-$0.53$\pm$0.10 &  $-$0.08$\pm$0.02 &   0.18$\pm$0.01 &   0.23$\pm$0.01 &   0.30$\pm$0.01 &   0.44$\pm$0.02 &  0.18$\pm$0.05 &  3.14$\pm$0.28 &  0.85$\pm$0.07 \\
%&  0.42$\pm$0.04 &  0.57$\pm$0.06 \\
13 &  53.054432 & $-$27.902388 &  $-$0.09$\pm$0.02 &   0.21$\pm$0.01 &   0.53$\pm$0.01 &  $-$0.39$\pm$0.08 &   0.35$\pm$0.01 &   0.43$\pm$0.01 &   0.31$\pm$0.03 & \nodata & \nodata &  0.39$\pm$0.05 &  9.06$\pm$0.22 &  1.85$\pm$0.07 \\
%&  0.46$\pm$0.07 &  0.65$\pm$0.10 \\
14 &  53.148323 & $-$27.901218 &  $-$0.18$\pm$0.03 &   0.24$\pm$0.02 &   0.39$\pm$0.02 &   0.46$\pm$0.14 &   0.47$\pm$0.02 &   0.43$\pm$0.01 &   0.56$\pm$0.02 &   0.58$\pm$0.02 &   0.76$\pm$0.03 &  0.68$\pm$0.08 &  3.51$\pm$0.45 &  1.23$\pm$0.13 \\
%&  1.31$\pm$0.06 &  1.17$\pm$0.08 \\
15 &  53.161755 & $-$27.897072 &  $-$0.26$\pm$0.02 &   0.68$\pm$0.04 &   1.13$\pm$0.04 &   1.75$\pm$0.16 &   1.23$\pm$0.07 &   1.18$\pm$0.02 &   1.61$\pm$0.04 &   1.65$\pm$0.03 &   1.97$\pm$0.04 &  1.59$\pm$0.15 &  7.15$\pm$0.70 &  2.61$\pm$0.23 \\
%&  0.38$\pm$0.05 &  2.12$\pm$0.06 \\
16 &  53.204742 & $-$27.894541 &  $-$0.16$\pm$0.03 &  $-$0.09$\pm$0.02 &   0.16$\pm$0.02 &  $-$0.66$\pm$0.12 &   0.24$\pm$0.02 &   0.19$\pm$0.01 &   0.31$\pm$0.02 &   0.32$\pm$0.02 &   0.46$\pm$0.02 &  0.34$\pm$0.06 &  2.46$\pm$0.36 &  0.84$\pm$0.10 \\
%& $-$0.08$\pm$0.04 &  0.65$\pm$0.06 \\
17 &  53.109936 & $-$27.880537 &  $-$0.15$\pm$0.02 &   0.09$\pm$0.02 &   0.16$\pm$0.02 &  $-$0.64$\pm$0.12 &   0.34$\pm$0.02 &   0.24$\pm$0.01 &   0.39$\pm$0.02 & \nodata & \nodata &  0.44$\pm$0.07 &  2.86$\pm$0.35 &  1.16$\pm$0.10 \\
%&  1.28$\pm$0.05 &  1.40$\pm$0.06 \\
18 &  53.171844 & $-$27.872406 &  $-$0.17$\pm$0.02 &   0.17$\pm$0.02 &   0.35$\pm$0.02 &  $-$0.70$\pm$0.13 &   0.40$\pm$0.02 &   0.37$\pm$0.01 &   0.45$\pm$0.02 &   0.38$\pm$0.02 &   0.45$\pm$0.03 &  0.44$\pm$0.08 &  3.67$\pm$0.40 &  0.93$\pm$0.11 \\
%&  0.60$\pm$0.05 &  0.37$\pm$0.07 \\
19 &  53.157902 & $-$27.869659 &  $-$0.28$\pm$0.04 &   0.54$\pm$0.03 &   0.77$\pm$0.03 &   1.44$\pm$0.20 &   1.03$\pm$0.03 &   0.89$\pm$0.02 &   0.70$\pm$0.03 &   0.84$\pm$0.04 &   1.47$\pm$0.04 &  0.75$\pm$0.13 &  4.27$\pm$0.67 &  1.55$\pm$0.21 \\
%&  1.17$\pm$0.05 &  1.06$\pm$0.07 \\
20 &  53.048820 & $-$27.865334 &  $-$0.27$\pm$0.04 &  $-$0.16$\pm$0.03 &  $-$0.19$\pm$0.04 &  $-$1.14$\pm$0.21 &  $-$0.17$\pm$0.03 &   0.07$\pm$0.02 &   0.26$\pm$0.08 &   0.38$\pm$0.05 &   0.80$\pm$0.06 &  0.50$\pm$0.12 &  3.30$\pm$0.65 &  1.54$\pm$0.17 \\
%&  4.12$\pm$0.07 &  4.93$\pm$0.10 \\
21 &  53.019222 & $-$27.847075 &  $-$0.10$\pm$0.02 &  $-$0.06$\pm$0.01 &   0.09$\pm$0.01 &  $-$0.44$\pm$0.08 &   0.18$\pm$0.01 &   0.13$\pm$0.01 &   0.24$\pm$0.03 & \nodata & \nodata &  0.46$\pm$0.04 &  3.31$\pm$0.24 &  1.54$\pm$0.07 \\
%&  1.91$\pm$0.07 &  2.27$\pm$0.13 \\
22 &  53.061619 & $-$27.846251 &  $-$0.20$\pm$0.03 &   0.72$\pm$0.02 &   1.24$\pm$0.03 &   1.26$\pm$0.15 &   1.30$\pm$0.02 &   1.27$\pm$0.01 &   1.39$\pm$0.03 &   1.11$\pm$0.02 &   1.29$\pm$0.03 &  0.98$\pm$0.15 &  4.36$\pm$0.61 &  1.91$\pm$0.18 \\
%&  1.30$\pm$0.06 &  1.26$\pm$0.09 \\
23 &  53.124718 & $-$27.824574 &  $-$0.15$\pm$0.02 &   0.42$\pm$0.01 &   0.77$\pm$0.02 &   0.94$\pm$0.11 &   0.93$\pm$0.02 &   0.85$\pm$0.01 &   1.17$\pm$0.02 &   1.11$\pm$0.01 &   1.39$\pm$0.02 &  0.99$\pm$0.06 &  6.25$\pm$0.39 &  2.29$\pm$0.09 \\
%&  2.11$\pm$0.04 &  2.17$\pm$0.05 \\
24 &  53.080879 & $-$27.791168 &  $-$0.14$\pm$0.02 &   0.15$\pm$0.02 &   0.36$\pm$0.02 &   0.48$\pm$0.12 &   0.46$\pm$0.02 &   0.41$\pm$0.01 &   0.62$\pm$0.02 &   0.66$\pm$0.01 &   0.88$\pm$0.02 &  0.53$\pm$0.05 &  5.22$\pm$0.39 &  1.57$\pm$0.09 \\
%&  1.44$\pm$0.05 &  1.51$\pm$0.08 \\
25 &  53.106331 & $-$27.783159 &  $-$0.15$\pm$0.02 &   0.14$\pm$0.02 &   0.18$\pm$0.02 &  $-$0.63$\pm$0.12 &   0.22$\pm$0.02 &   0.20$\pm$0.01 &   0.29$\pm$0.02 &   0.43$\pm$0.01 &   0.62$\pm$0.02 &  0.45$\pm$0.05 &  2.83$\pm$0.36 &  1.16$\pm$0.09 \\
%&  1.17$\pm$0.04 &  1.16$\pm$0.06 \\
26 &  53.263027 & $-$27.759014 &  $-$0.16$\pm$0.03 &   0.26$\pm$0.02 &   0.43$\pm$0.02 &   0.51$\pm$0.13 &   0.60$\pm$0.02 &   0.51$\pm$0.01 &   0.56$\pm$0.05 & \nodata & \nodata &  0.50$\pm$0.10 &  4.09$\pm$0.40 &  1.21$\pm$0.13 \\
%&  1.00$\pm$0.08 &  1.13$\pm$0.10 \\
27 &  53.008801 & $-$27.758272 &  $-$0.16$\pm$0.03 &   0.10$\pm$0.02 &   0.19$\pm$0.02 &  $-$0.69$\pm$0.13 &   0.29$\pm$0.02 &   0.23$\pm$0.01 &   0.31$\pm$0.03 & \nodata & \nodata &  0.31$\pm$0.07 &  3.93$\pm$0.37 &  1.42$\pm$0.11 \\
%&  0.19$\pm$0.11 &  0.19$\pm$0.13 \\
28 &  53.013573 & $-$27.755177 &  $-$0.14$\pm$0.02 &   0.39$\pm$0.02 &   0.89$\pm$0.02 &   0.56$\pm$0.12 &   0.66$\pm$0.02 &   0.76$\pm$0.01 &   0.65$\pm$0.02 &   0.66$\pm$0.01 &   0.84$\pm$0.02 &  0.62$\pm$0.05 & 10.49$\pm$0.33 &  2.93$\pm$0.08 \\
%&  0.59$\pm$0.07 &  0.85$\pm$0.11 \\
29 &  53.110451 & $-$27.754616 &  $-$0.12$\pm$0.02 &  $-$0.07$\pm$0.01 &  $-$0.09$\pm$0.02 &  $-$0.53$\pm$0.10 &  $-$0.08$\pm$0.02 &   0.23$\pm$0.01 &   0.41$\pm$0.01 &   0.52$\pm$0.01 &   0.81$\pm$0.01 &  0.46$\pm$0.05 &  2.91$\pm$0.31 &  1.29$\pm$0.08 \\
%&  0.98$\pm$0.06 &  1.91$\pm$0.07 \\
30 &  53.140198 & $-$27.751116 &  $-$0.18$\pm$0.03 &   0.18$\pm$0.02 &   0.28$\pm$0.02 &  $-$0.77$\pm$0.14 &   0.32$\pm$0.02 &   0.30$\pm$0.01 &   0.39$\pm$0.02 &   0.29$\pm$0.02 &   0.38$\pm$0.02 &  0.25$\pm$0.08 &  2.57$\pm$0.48 &  0.61$\pm$0.13 \\
%&  0.50$\pm$0.05 &  0.68$\pm$0.07 \\
31 &  53.131378 & $-$27.745434 &  $-$0.19$\pm$0.03 &   0.16$\pm$0.02 &   0.33$\pm$0.02 &  $-$0.79$\pm$0.14 &   0.41$\pm$0.02 &   0.37$\pm$0.01 &   0.54$\pm$0.02 &   0.56$\pm$0.02 &   0.68$\pm$0.02 &  0.46$\pm$0.08 &  3.65$\pm$0.48 &  1.53$\pm$0.12 \\
%&  1.48$\pm$0.05 &  1.29$\pm$0.07 \\
32 &  53.124756 & $-$27.744980 &  $-$0.15$\pm$0.02 &   0.10$\pm$0.02 &   0.18$\pm$0.02 &  $-$0.62$\pm$0.12 &   0.26$\pm$0.02 &   0.22$\pm$0.01 &   0.32$\pm$0.02 &   0.24$\pm$0.01 &   0.38$\pm$0.02 &  0.40$\pm$0.07 &  4.13$\pm$0.39 &  1.01$\pm$0.11 \\
%&  0.48$\pm$0.05 &  0.61$\pm$0.09 \\
33 &  53.214493 & $-$27.739864 & \nodata & \nodata & \nodata & \nodata & \nodata &   0.09$\pm$0.02 &   0.57$\pm$0.06 & \nodata & \nodata &  0.62$\pm$0.08 &  6.92$\pm$0.36 &  3.65$\pm$0.14 \\
%& 11.72$\pm$0.14 & 16.40$\pm$0.21 \\
34 &  53.087608 & $-$27.726042 &  $-$0.17$\pm$0.05 &   0.38$\pm$0.02 &   0.52$\pm$0.02 &   0.72$\pm$0.14 &   0.59$\pm$0.02 &   0.55$\pm$0.01 &   0.77$\pm$0.02 &   0.85$\pm$0.01 &   1.04$\pm$0.01 &  0.69$\pm$0.06 &  2.25$\pm$0.04 &  1.08$\pm$0.10 \\
%& 1.42$\pm$0.04 &  1.51$\pm$0.07 \\
\enddata
%\vskip -3in
\vspace{-1in}
\end{deluxetable*}
{\hskip -3.5in Note that all fluxes are given in units of $\mu$Jy.}
\end{longrotatetable}

The redshift distribution of the 34 \OIII\ ELGs and H$_2\mathrm{S}(1)$ width are shown in Figure~\ref{f:zdistribution.pdf}.  Because the narrowband filter H$_2\mathrm{S}(1)$ has a relatively small band width $\Delta\lambda$, we attribute the emission line detected in this band to \OIII$\lambda$5007 (this will be discussed later in Section~\ref{sec:bias}).  More details about emission-line source selection and the EAZY SED fittings can ben found in \citet{An2013} and \citetalias{An2014}.

\section{SED Fitting} \label{sec:sedfitting}

In order to maximize signal-to-noise ratio (S/N) for aperture-matched colors between the 12 bands mentioned above, \citetalias{An2014} firstly determined colors for MUSYC, \textit{HST} and CFHT bands, respectively, and then matched three sets of colors to establish SED input fluxes from $U$ to $K_\mathrm{s}$.  Finally, the SED was scaled up to meet the total flux of $K_\mathrm{s}$ derived from aperture photometry within a diameter of 2$\arcmin$ corrected for the missing flux out of the aperture \citepalias[see][for more details]{An2014}.  The photometric fluxes in the 12 bands from $U$ to $K_\mathrm{s}$ for our sample of 34 \OIII\ ELGs are listed in Table~\ref{tab:photometry}.

Photometric redshifts (photo-$z$) were derived from these SEDs using EAZY with relatively small errors partially because of the narrowband data points linked to given redshifts traced by emission lines.  The galaxy SED templates were generated from a library of six independent templates in EAZY.  This gives a fast determination on photo-$z$ but sacrifices the accuracy in modeling the details of SEDs (e.g., line fluxes). Therefore, we utilize the Code Investigating GALaxy Emission  \citep[CIGALE;][]{Boquien2019} to analyze the SEDs of our \OIII\ ELGs with the improved galaxy templates.  CIGALE produces millions of models to fit the observational data and estimates their physical properties such as stellar mass, star formation rate (SFR), and dust attenuation, while applying a Bayesian statistical analysis approach to estimate the results.  The photo-$z$ obtained with EAZY are used in the CIGALE fitting as the input redshift since CIGALE is not optimized to measure photometric redshift.

The \citet{Chabrier2003} IMF and the stellar population synthesis model from \citet{BC2003} are adopted for the fitting.  We set three values of metallicity as 0.0004, 0.008 and 0.02 (for Z$_\odot$=0.02) in stellar models.  A delayed form of star formation history (SFH), SFR$\,\propto t/\tau_\mathrm{main}^2\,\exp(-t/\tau_\mathrm{main})$, is adopted in our fitting too, where $t$ is the time of the star-formation onset and $\tau_\mathrm{main}$ is the $e$-folding time of the main stellar population.  Such a functional form is more physical than a simple exponential SFH because it removes the discontinuity in SFR at $t=0$ and is able to produce an increasing SFR when $\tau$ is large \citep{Carnall2019}.  A starburst component f$_\mathrm{burst}$ can be added at a given mass fraction as well.  This SFH can fit the high-$z$ SFGs well since they usually have a relatively strong star formation activity.  It also avoids the systematic biases caused by the degeneracy between the slope of the dust attenuation curve, effective dust attenuation, and the intrinsic UV slope of model templates \citep{Yuan2019,Villa-Velez2021,Qin2022}.  The degeneracy is indeed smaller for blue SFGs. However, high-$z$ SFGs may have nonnegligible dust attenuation, and this could increase the degeneracy for these dusty objects.

For nebular emission, the initial parameters include the dimensionless ionization parameter $\log U$, the escape fraction of Lyman continuum photons $f_\mathrm{esc}$ and the fraction of Lyman continuum photons absorbed by dust $f_\mathrm{dust}$.  The radiation strength $U$ is set to be $-2.0$, while $f_\mathrm{esc}$ and $f_\mathrm{dust}$ have values of 0.0, 0.1, and 0.2.

We adopt the modified Calzetti law \citep{Calzetti2000,Noll2009b} to describe the dust attenuation in our fitting.  The attenuation for nebular emission is higher than that for stellar emission.  We set $E(B-V)_\mathrm{factor}$, a ratio of $E(B-V)_\mathrm{star}$ to the $E(B-V)_\mathrm{gas}$, to be 0.44 \citep{Calzetti2000}.  Meanwhile, the color excess of the nebular lines $E(B-V)_\mathrm{gas}$ and the slope of the power law modifying the attenuation curve $\delta$ are chosen to change freely, in the range of 0.05--0.8 and $-1.0$ to $0.2$, respectively.  We also take the amplitude of UV bump to be 0, 1, 2, 3, or 4, where 3 corresponds to the value of the Milky Way.  And the ratio of total to selective extinction $R_v$ is fixed to a standard value of 3.1.

Due to the lack of detections in the far-IR, the SED of dust emission is not well constrained in our work.   The dust emission is modeled with templates from \citet{Dale2014}, which refines the PAH emission and also adds an optional active galactic nucleus (AGN) fraction in the modeling.  The star-forming component is parameterized by a single parameter $\alpha$ defined as d$M_{\rm d}(U)\propto U^{-\alpha}{\rm d}U$, where $M_{\rm d}$ is the dust mass and $U$ is the radiation field intensity.  We set the AGN fraction to be zero, and the slope is fixed at $\alpha=2$.

 As pointed out by \citet{Lambrides2020}, the X-ray selection might miss faint AGNs at $z>2$ because their host galaxies usually have strong star formation activities.  As a check, we add detailed AGN models from \citet{Fritz2006} in CIGALE to test our results.  The ratio of the maximum to minimum radii of the dust torus is set to be 30 and 100, while the optical depth at 9.7\,$\micron$ is 0.3 and 2.  And the AGN fraction has a value of 0, 0.01, and 0.1.  All the input parameters of CIGALE fitting are given in Table~\ref{tab:cigale}.

\begin{figure*}[tbp]
\centering
\includegraphics[width=0.96\textwidth]{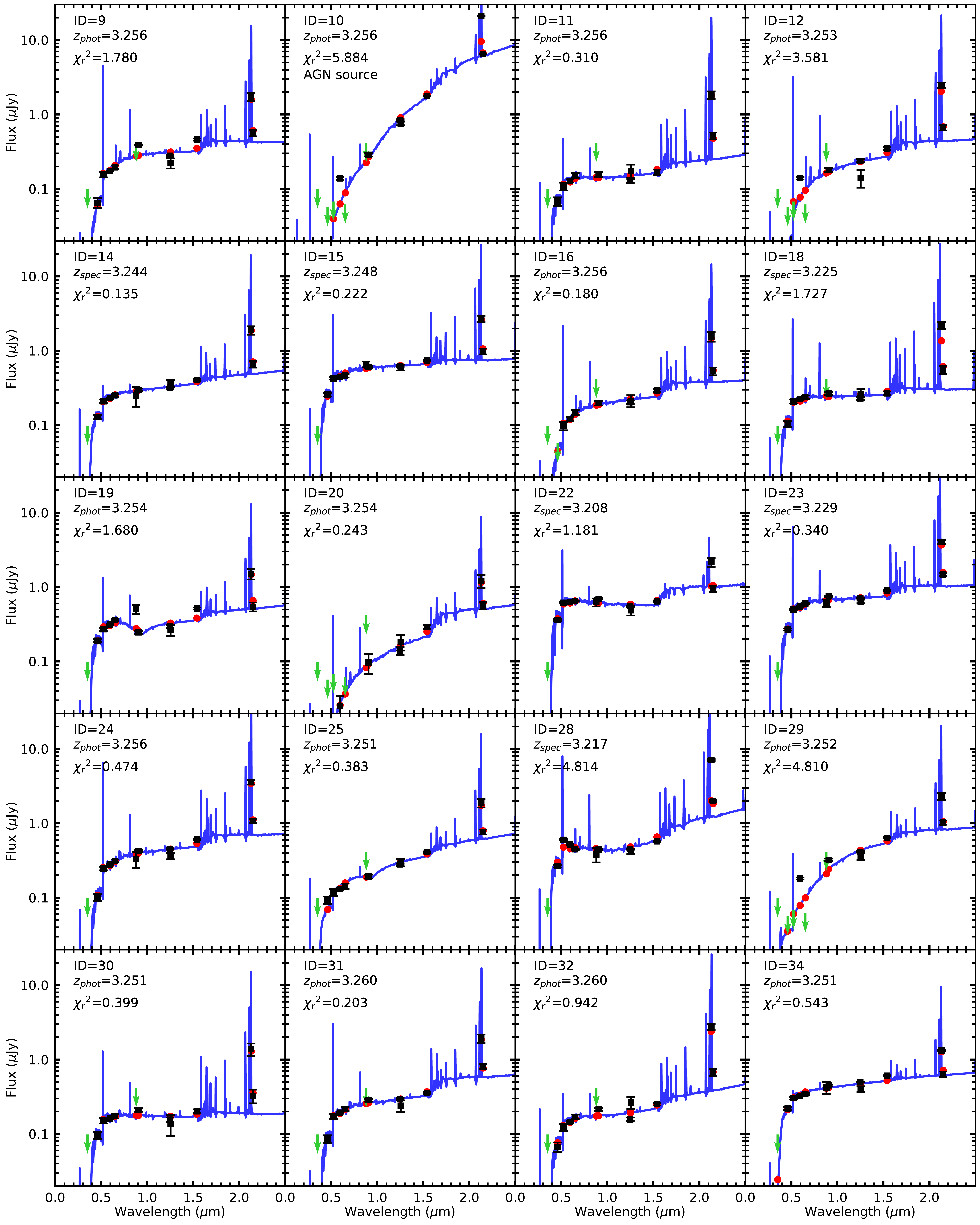}
\caption{Best-fit SEDs from CIGALE for our sample of 34  $z\sim3.25$ \OIII\ ELGs. 
		Black squares and error bars present the observed data points of multiband photometry. 
		Green arrows mark the upper limits of given band detections. 
		Blue lines are the best-fit model SEDs, and red circles refer to the model points in the same bands. 
		The ID, photo-$z$ (spec-$z$ if available), and $\chi_r^2$ are labeled in each panel.
		The object with ID=10 hosts an AGN detected in the 7\,Ms \textit{Chandra} observations.
		These 20 \OIII\ ELGs have \textit{HST}/WFC3 $J_\mathrm{125}$ and $H_\mathrm{160}$ imaging data.}
\label{f:sed.pdf}
\end{figure*}

\begin{figure*}[tbp]
\centering
\figurenum{2}
\includegraphics[width=0.96\textwidth]{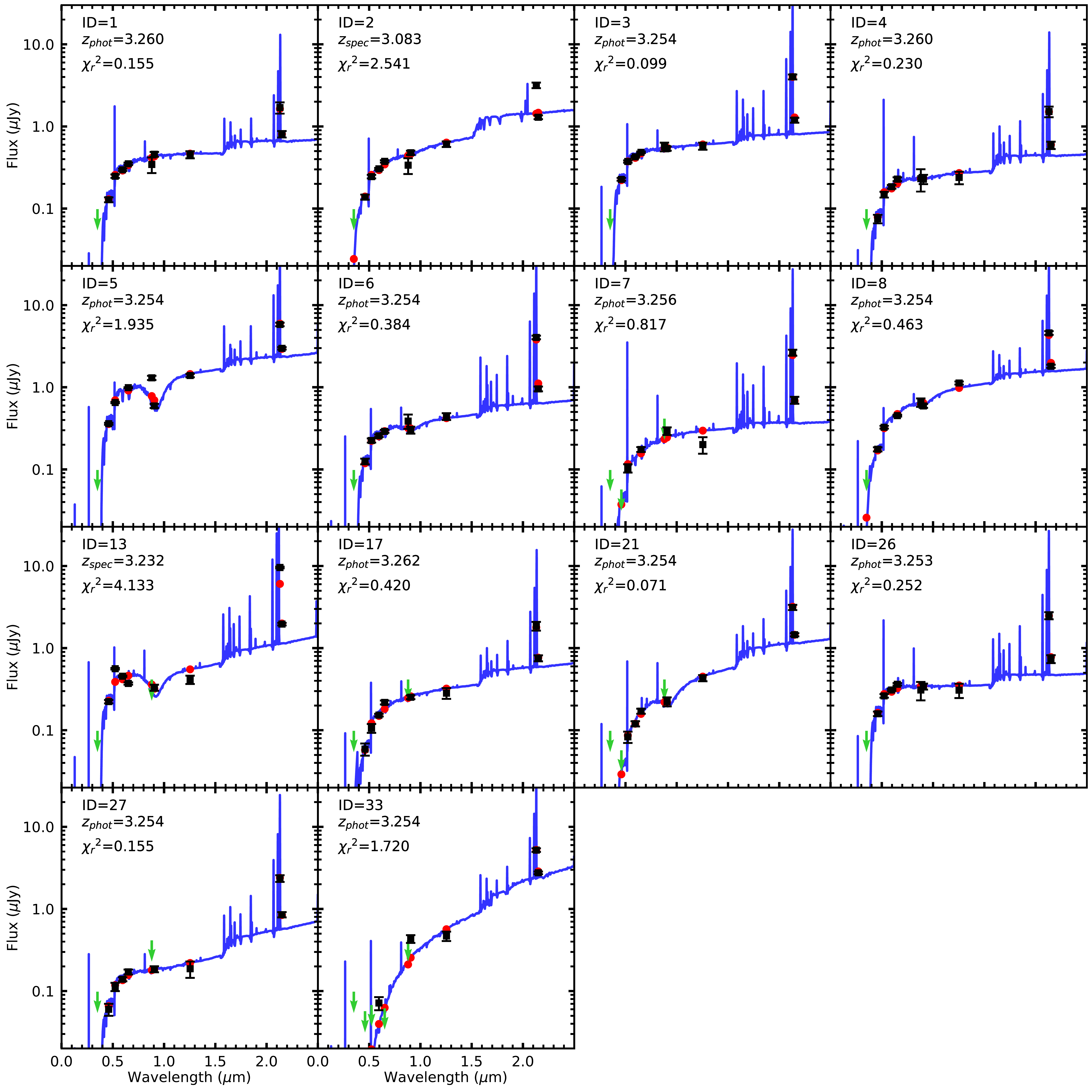}
\caption{Continued for the best-fit SEDs from CIGALE fitting of the remaining 14 \OIII\ ELGs, which are not covered 
		by the \textit{HST}/WFC3 $J_\mathrm{125}$ and $H_\mathrm{160}$ imaging.}
\end{figure*}

The best-fit model from CIGALE is chosen using the least-squares method, and the reduced $\chi^{2}$ ($\chi_\mathrm{r}^2=\chi^{2}/(N-1)$) is used as a global indicator to quantify the quality of the fitting.  Figure~\ref{f:sed.pdf} shows the results of CIGALE fitting of 34 \OIII\ ELGs; their ID, photo-$z$ (spec-$z$ if available), and parameter $\chi_\mathrm{r}^2$ of the fittings are shown in each panel.  The first figure consists of 20 \OIII\ ELGs with $J_\mathrm{125}$ and $H_\mathrm{160}$ imaging data, while the second panel consists of the remaining 14 \OIII\ ELGs without these two bands.  Only one object (ID=10) shows a large $\chi_\mathrm{r}^2>5$ owing to the detection upper limits in five MUSYC bands.  About 65\% of the objects have $\chi_\mathrm{r}^2<1$ and 85\% have $\chi_\mathrm{r}^2<3$, suggesting that most of our best-fit SEDs are obtained in good quality.

Note that the 2175\AA\ bump is present in the best-fit SEDs for some \OIII\ ELGs in our sample (e.g., ID=5, 13).  We will further examine in Section~\ref{sec:attenuaion} whether such a feature is caused by measurement uncertainties or a solid detection of the 2175\AA\ bump in $z\sim 3.25$ \OIII\ ELGs.

\begin{table*}
\caption{Input Parameters for SED fitting with CIGALE \citep{Boquien2019}  \label{tab:cigale}}
 
\scalebox{0.82}
{
\begin{tabular}{c c c}     % 3 columns 
\hline\hline       
                      % To combine 4 columns into a single one 
\textbf{Parameter} & \textbf{Symbol} & \textbf{Value}\\ 
\hline                    
    & \textbf{Stellar Population} \\
    & \citet{BC2003} & \\ %\hline
Initial mass function   & IMF & \cite{Chabrier2003} \\  
Metallicity    & Z$_\mathrm{\star}$ & 0.0004, 0.008, 0.02 \\ \hline
    & \textbf{Delayed Star Formation History} &  \\ %\hline
$e$-folding time of main stellar population~(Myr) &  $\tau_\mathrm{main}$ & 50, 100, 200, 500, 4000, 10000 \\
Age of main stellar population~(Myr)    & age$_\mathrm{main}$ & 50, 500, 1000, 2000 \\  
Age of the late burst~(Myr)    & age$_\mathrm{burst}$ & 20,50 \\
$e$-folding time of the late burst~(Myr) & $\tau_\mathrm{burst}$ & 10, 50, 100, 1000, 8000\\
Mass fraction of the late burst population    & f$_\mathrm{burst}$ & 0.01, 0.1, 0.2 \\  \hline
    & \textbf{Nebular Emission} &  \\  %\hline
Ionization parameter   & $\log U$ & $-$2.0 \\
LyC escape fraction    & $f_\mathrm{esc}$   & 0.0, 0.1, 0.2 \\
LyC absorbed by dust   & $f_\mathrm{dust}$  & 0.0, 0.1, 0.2 \\ \hline
    & \textbf{Dust attenuation} & \\
    & \citet{Noll2009b} &\\ %\hline
Color excess of the nebular lines~(Mag) & $E(B-V)_\mathrm{lines}$ & 0.05, 0.1, 0.15, 0.2, 0.25, 0.3, 0.4, 0.5, 0.6, 0.7, 0.8 \\  
Nebular-to-continuum ratio   & $E(B-V)_\mathrm{factor}$ &  0.44\\
Amplitude of the UV bump   & $A_\mathrm{bump}$ &  0, 1, 2, 3, 4\\
Slope of the attenuation curve   & $\delta$ & $-$1.0, $-$0.8, $-$0.6, $-$0.4, $-$0.2, 0.0, 0.2 \\  
Line emission extinction    & $R_v$ &  3.1\\ \hline   
    & \textbf{Dust Emission} & \\
    & \citet{Dale2014} &\\ %\hline
Power-law slope $\mathrm{dU/dM}\propto\mathrm{U}^\alpha$    & $\alpha$ & 2.0 \\ \hline
    & \textbf{AGN Model} & \\
    & \citet{Fritz2006} &\\ %\hline
Ratio of the max to min radii of the dust torus & r$_{\rm ratio}$ & 30, 100 \\
Torus optical depth at 9.7\,$\micron$ & $\tau_{9.7}$ & 0.3, 2.0 \\
Density function parameter beta & $\beta$ & -0.5 \\
Density function parameter gamma & $\gamma$ & 4.0 \\
Opening angle of the dust torus~($^{\circ}$) & $\theta$ & 100 \\
Viewing angle~($^{\circ}$) & $\psi$ & 30.1, 60.1 \\
AGN fraction & $f_\mathrm{frac}$ & 0.0, 0.01, 0.1 \\
\hline                  
\end{tabular}
}
\end{table*}

\section{Properties of \OIII\ Emission-line Galaxies} \label{sec:results}

\subsection{X-ray Detection} \label{sec:X-ray}

AGN activities can also make contribution to the \OIII\ line.  Thus, it is essential to verify whether the \OIII\ emission lines in some of our sample galaxies are powered by AGNs.  We firstly match the X-ray source catalog of 7\,Ms \textit{Chandra} observations in CDFS from \citet{Luo2017} with our sample of \OIII\ ELGs to examine their X-ray properties.  Of the 34 \OIII\ ELGs, only one object is identified as an X-ray source (XID=760) in the 7\,Ms catalog.  Its absorption-corrected intrinsic 0.5$-$7.0\,keV luminosity is $1.12\times10^{45}$\,erg\,s$^{-1}$.  Moreover, this object has the highest \OIII\ luminosity $L_\mathrm{\OIII}=10^{44.2}$\,erg\,s$^{-1}$, as well as highest stellar mass $M_\ast=10^{11.4}\,\Msun$ (see Table~\ref{tab:properties}).  No other X-ray counterparts are found in the 7\,Ms catalog for the remaining \OIII\ ELGs in our sample.   The CIGALE results also show that the majority of our \OIII\ ELGs contain no or a negligible AGN component.  Including the X-ray source, six sample galaxies in our sample have an AGN fraction of 0.1, while about half have no AGN fraction in the fitting.

\subsection{Morphologies} \label{sec:morphology}

We use \textit{HST}/ACS $V_\mathrm{606}$ and $z_\mathrm{850}$ imaging data from the GOODS and GEMS surveys to investigate the morphologies of our sample galaxies.  For $z\sim3.25$, these two bands correspond to the rest-frame far-UV (FUV: 1402\AA) and near-UV (NUV: 2130\AA).  Note that three \OIII\ ELGs  do not have $V_\mathrm{606}$ imaging data owing to the incomplete coverage of the GEMS observations.  We adopt the \textit{HST}/ACS F814W ($I_\mathrm{814}$) imaging data to replace $V_\mathrm{606}$ for making color images of these three objects.

\begin{figure*}
\centering
\includegraphics[width=0.96\textwidth]{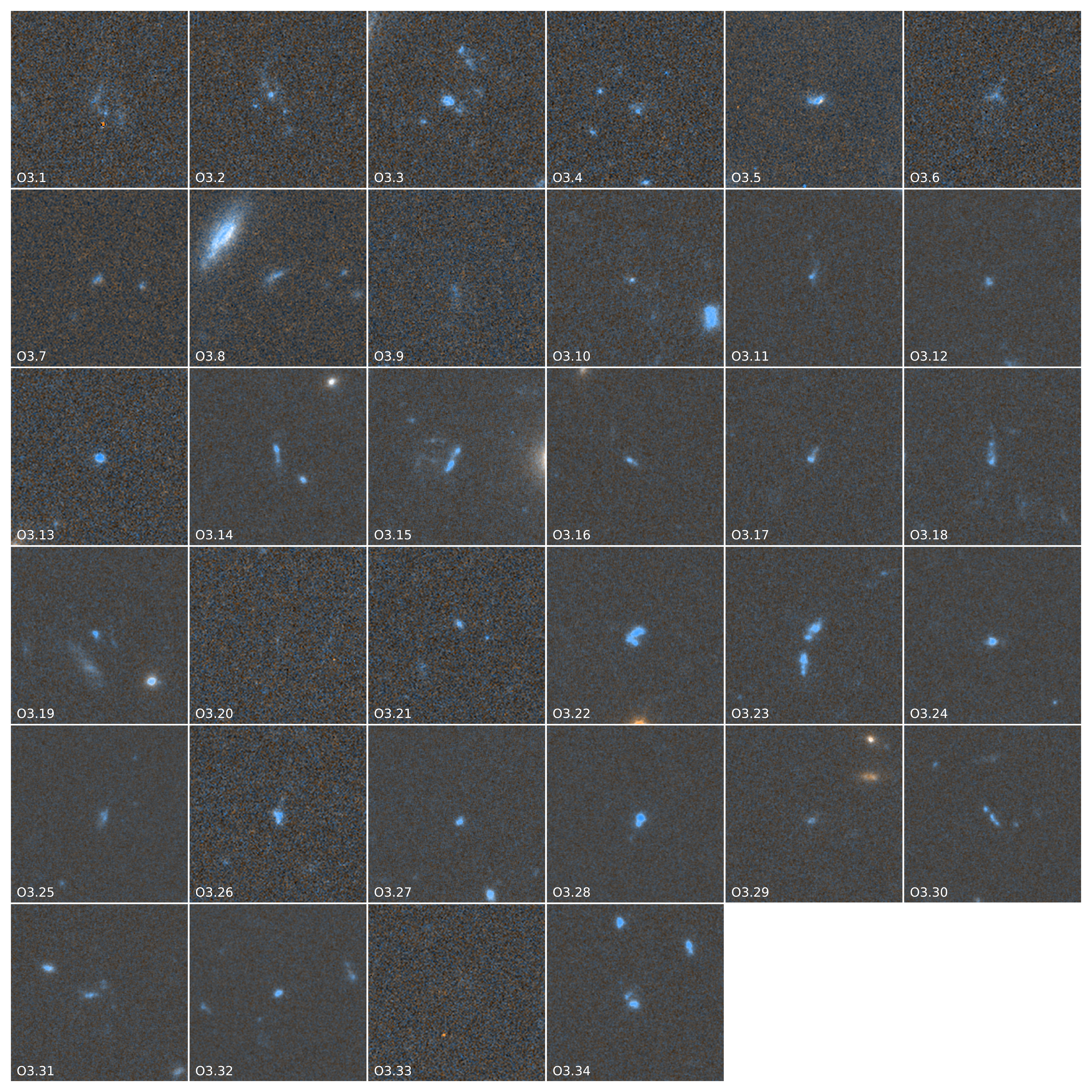}
\caption{\textit{HST}/ACS stamp images for our sample of 34 \OIII\ ELGs. 
		Each stamp is given in a size of $6\arcsec \times 6\arcsec$, 
		corresponding to 45\,kpc\,$\times$\,45\,kpc at $z\sim3.25$. 
		These color images are made with \textit{HST}/ACS $V_\mathrm{606}$ and $z_\mathrm{850}$ images 
		from the GEMS and CANDELS surveys. 
		Note that three images (ID=6, 7, 8) are made with $I_\mathrm{814}$ and $z_\mathrm{850}$ images owing to the lack of 
		the $V_\mathrm{606}$ observations. }
\label{f:ACS.pdf}
\end{figure*}

Figure~\ref{f:ACS.pdf} shows the color images made with $V_\mathrm{606}$ and $z_\mathrm{850}$ for our \OIII\ ELGs.  A variety of morphologies can be seen.   Here we define five morphological types in terms of the compactness and numbers of components of a galaxy.  We assign the types of UV faint, compact, diffuse (including clumpy and tidal ones), merging, and multiple component with type ID 1--5, respectively.  The morphological types of these 34 \OIII\ ELGs are visually classified by three of us (R.W., J.R. and S.L.).  The median of three classifications is adopted for each galaxy, and the results are presented in Table~\ref{tab:properties}.

Our results of morphological classifications show that about 6\% (2/34) are too faint to be securely resolved in both $V_\mathrm{606}$ and $z_\mathrm{850}$; 35\% (12/34) appear to be compact with $R_\mathrm{e}<0\farcs3$ and the majority of these compact ones are relatively bright; 38\% (13/34) of our sample \OIII\ ELGs have diffuse emission out to $R=1\arcsec-2\arcsec$ (7--15\,kpc) or have apparent tidal/clumpy features within 1$\arcsec$;  9\% (3/34) look like mergers, with two obvious galactic nuclei connected by tidal bridges; and 12\% (4/34) are pairs of two or three components with comparable colors and sizes and are separated by $<2\arcsec$ without clear tidal bridges between them.

\begin{figure*}
\centering
\includegraphics[width=0.96\textwidth]{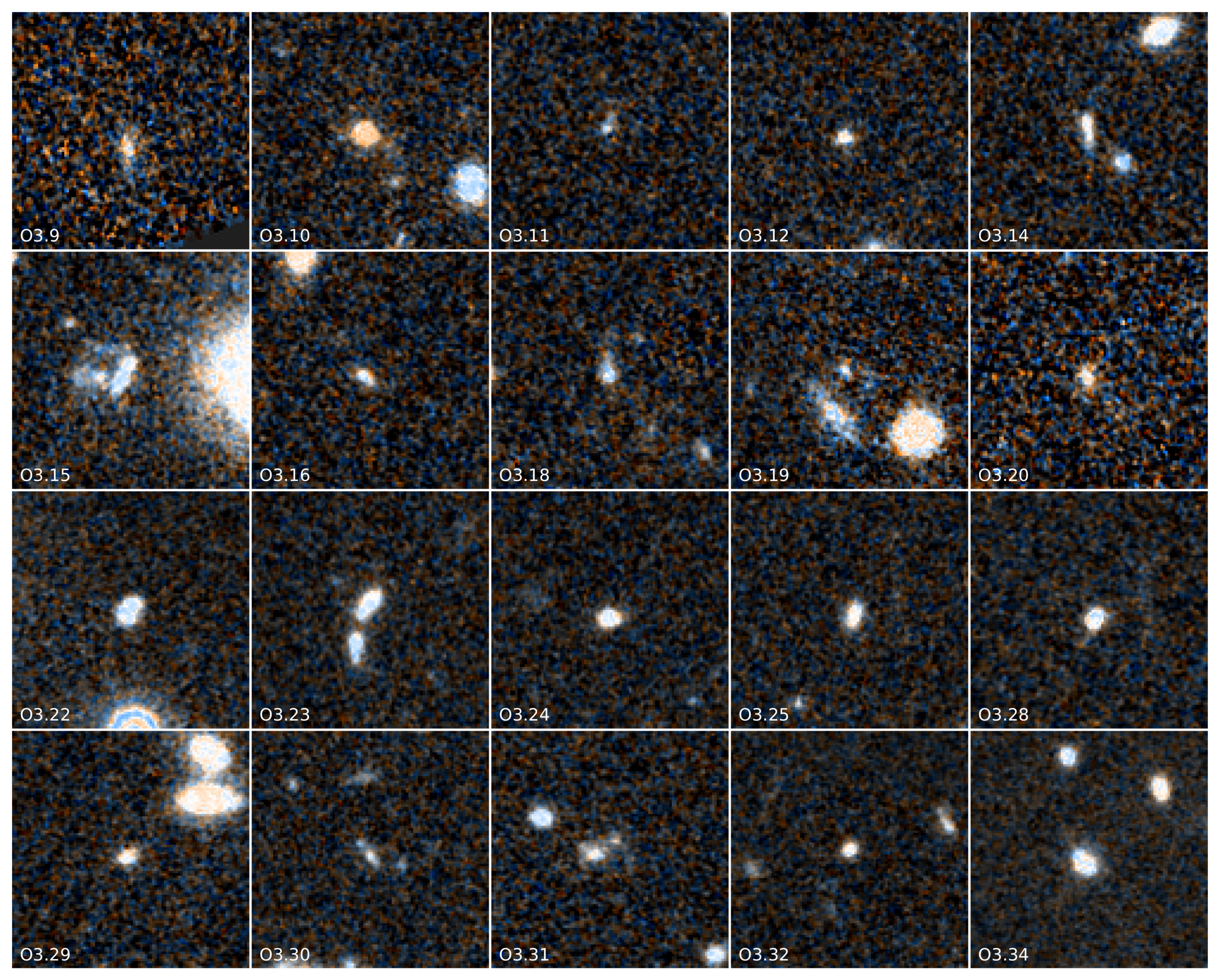}
\caption{\textit{HST}/WFC3 stamp images of 20 \OIII\ ELGs in our sample. 
		The color images are made with \textit{HST}/WFC3 $J_\mathrm{125}$ and $H_\mathrm{160}$ images from CANDELS (the same size as given in Figure~\ref{f:ACS.pdf}).}
\label{f:WFC3.pdf}
\end{figure*}

There are 20 sample ELGs having the \textit{HST}/WFC3 $J_\mathrm{125}$ and $H_\mathrm{160}$ images from CANDELS.  For $z\sim3.25$, these two bands correspond to the rest-frame NUV (2939\AA) and U (3624\AA).  Figure~\ref{f:WFC3.pdf} presents their color images made with the two-band data.  We carry out morphological classification with the $J_\mathrm{125}$ and $H_\mathrm{160}$ images.  The results are very similar to those based on the $V_\mathrm{606}+z_\mathrm{850}$ images.  Note that one object (ID=20) is invisible in $V_\mathrm{606}$and $z_\mathrm{850}$ and appears as a compact galaxy in $J_\mathrm{125}$ and $H_\mathrm{160}$.  We take the morphological classifications based on $V_\mathrm{606}$ and $z_\mathrm{850}$ as our main morphology properties for these 34 \OIII\ ELGs.

We point out that there are two UV-faint \OIII\ ELGs (ID=20, 33) being heavily attenuated by dust in rest-frame UV.  As shown in Figure~\ref{f:sed.pdf}, these two objects' SEDs show clear dust reddening, while several other objects (e.g., ID=10, 12, 29) also exhibit reddening from their SEDs.  We find that these galaxies are all compact, indicating that the compactness helps to maintain a dustier star-forming environment, although the rest-UV morphologies are very sensitive to young stellar populations with large uncertainties due to dust attenuation.   We stress that our sample selection based on the H$_2\mathrm{S}(1)$ and $K_\mathrm{s}$ observations is able to pick these dusty \OIII\  ELGs, which were missed by the previous studies based on the optical sample selections.

\subsection{Colors and \OIII\ EWs} \label{sec:rest-frame}

\begin{figure}
\centering
\includegraphics[width=0.45\textwidth]{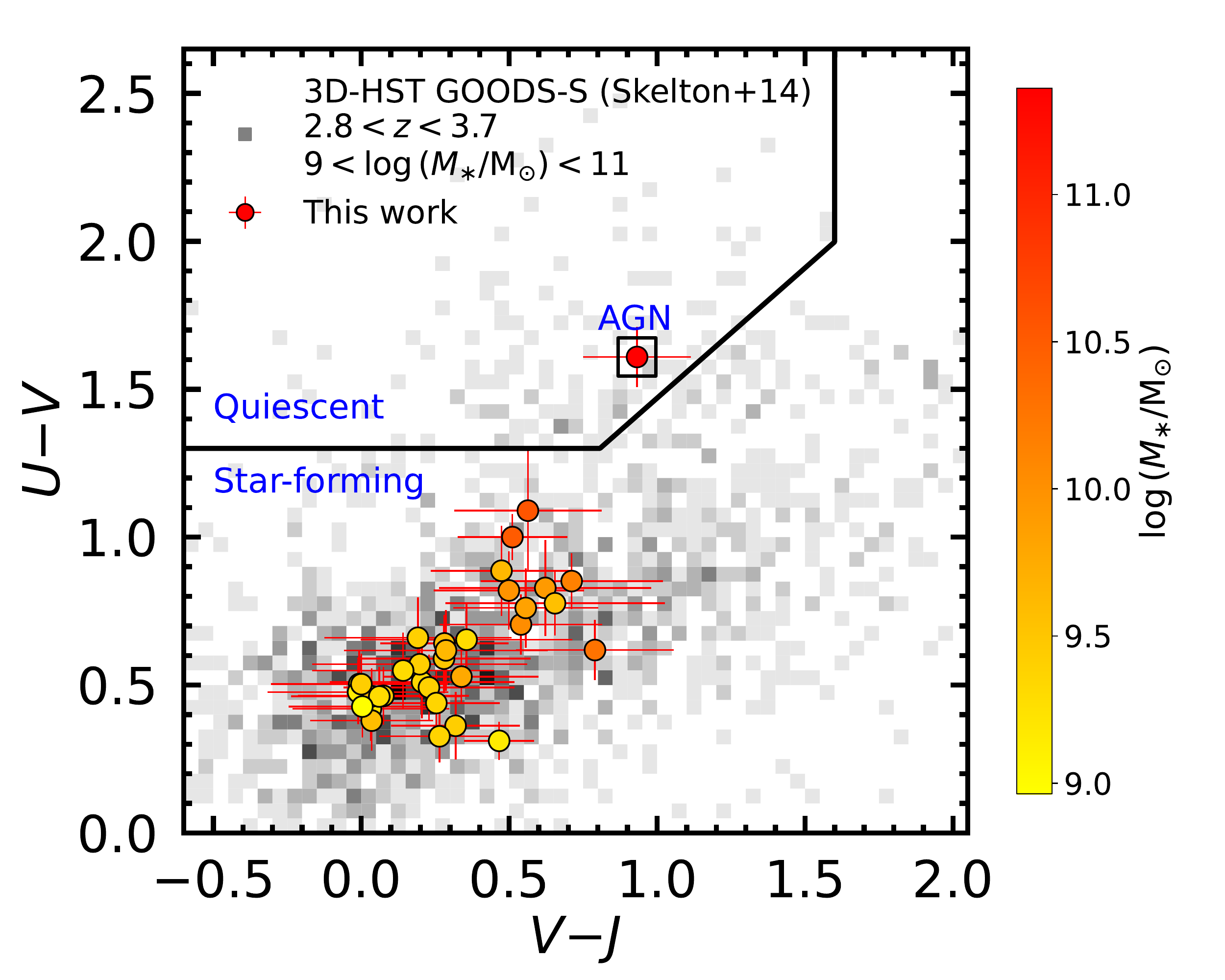}
\caption{Rest-frame $UVJ$ color diagram of 34 \OIII\ ELGs color-coded with stellar mass.
	The gray density map shows the distribution of  galaxies selected with $9<\log\,(M_{\ast}/$M$_{\odot})<11$ at $2.8<z<3.7$ from the 3D-HST GOODS-South catalog. 
	The selection box to separate star-forming and quiescent galaxies is from \citet{Williams2009}. 
	The AGN source in the regime of quiescent galaxies is marked with a black square. 
	Our sample \OIII\ ELGs are mostly located at the blue end of the star-forming regime. }
\label{f:UVJ.pdf}
\end{figure}

The $UVJ$ diagram is widely used to separate SFGs and quiescent galaxies \citep{Williams2009, Brammer2011, Whitaker2011}.  Similarly, the $U$ and $V$ filters given in \citet{Apellaniz2006}, and the $J$ filter from the Two Micron All Sky Survey (2MASS) are adopted to derive the rest-frame $U-V$ and $V-J$ colors from the best-fit model SEDs of our \OIII\ ELGs.  The calculation is done with CIGALE \citep{Boquien2019}.  Figure~\ref{f:UVJ.pdf} shows the distribution of our \OIII\ ELGs in the rest-frame $U-V$ and $V-J$ diagrams. The stellar masses derived from CIGALE are used to color-code the data points.  The selection criteria to distinguish star-forming and quiescent galaxies are adopted from \cite{Williams2009}.  For a comparison, we also show a sample of galaxies with $9<\log\,(M_{\ast}/$M$_{\odot})<11$ at $2.8<z<3.7$ from the 3D-HST GOODS-South catalog \citep{Skelton2014}.

It is clear from Figure~\ref{f:UVJ.pdf} that all of our sample \OIII\ ELGs are located in the star-forming regime, except for the AGN host with $U-V=1.6$ owing to strong dust attenuation.  This confirms that the vast majority of our \OIII\ sample galaxies are SFGs.  About 56\% (19/34) of the \OIII\ ELGs with $V-J<0.3$ are very blue, likely being little affected by dust attenuation or boosted in $V$ by strong \OIII\ emission.  Meanwhile, the \OIII\ ELGs of lower stellar masses tend to have bluer colors in both $U-V$ and $V-J$, consistent with the results for the overall galaxy population \citep{Skelton2014, Straatman2016}.

As described in Section~\ref{sec:data} \citepalias[see also][]{An2014}, the rest-frame EWs of our \OIII\ samples are estimated using the formula from \citet{Geach2008} as 
\begin{equation} \label{EW}
\mathrm{EW}_{\mathrm{\OIII}} = \frac{\Delta\lambda_{H_2\mathrm{S}(1)}\times (f_{H_2\mathrm{S}(1)} - f_{K_\mathrm{s}})}{[f_{K_\mathrm{s}} - f_{H_2\mathrm{S}(1)}(\Delta\lambda_{H_2\mathrm{S}(1)} /  \Delta\lambda_{K{_\mathrm{s}}})]\times(1+z)},
\end{equation}
where $\Delta\lambda_{H_2\mathrm{S}(1)}$ and $\Delta\lambda_{K_{\mathrm{s}}}$ are the widths of the narrow- and broadband filters and $f_{H_2\mathrm{S}(1)}$ and $f_{K_\mathrm{s}}$ are the flux densities in these two bands.  The estimated EWs are listed in Table~\ref{tab:properties}.

\begin{figure}
\centering
\includegraphics[width=0.45\textwidth]{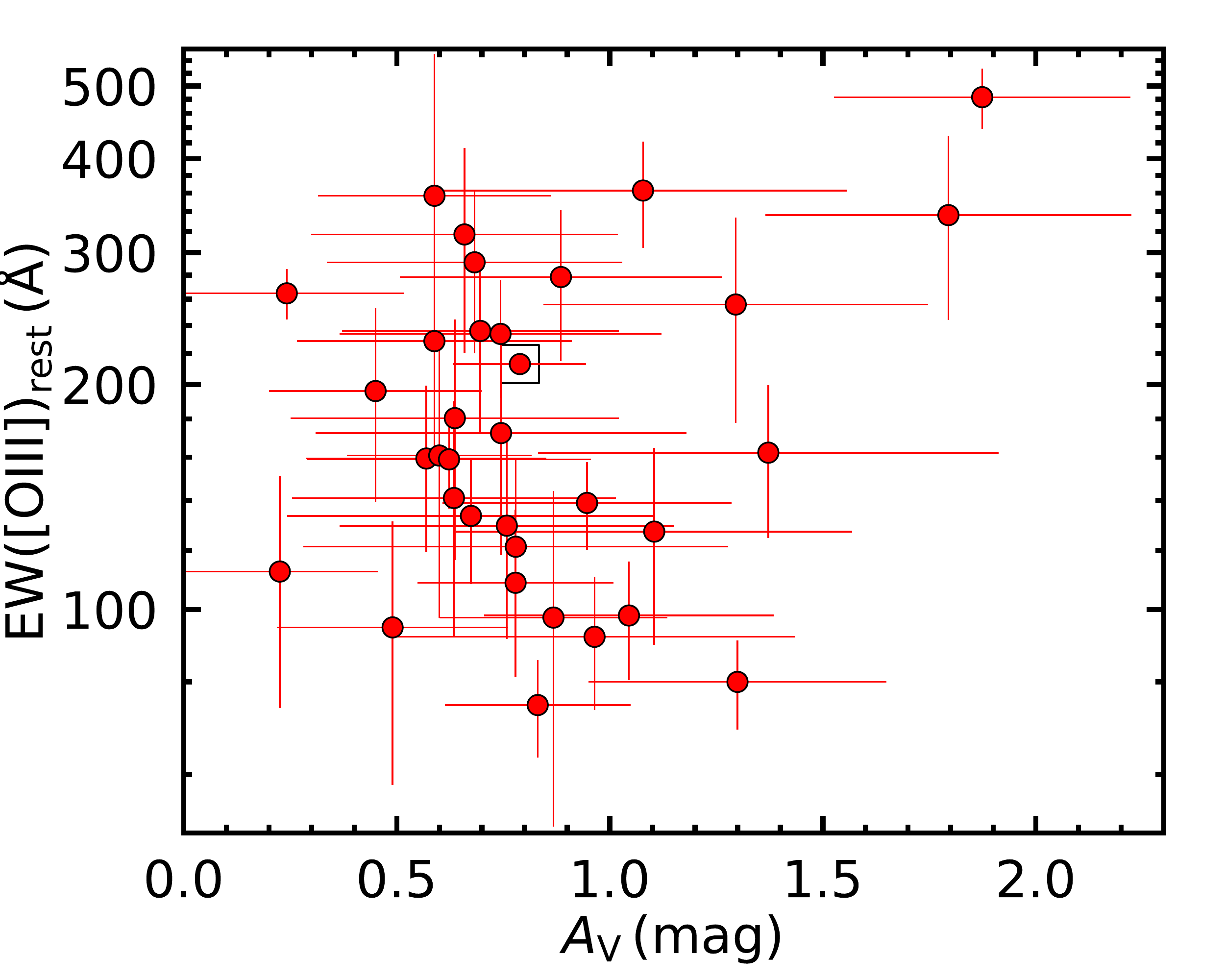}
\caption{$A_V$ versus rest-frame \OIII\ EW  for our sample of \OIII\ ELGs. 
    The majority of our sample galaxies have rest-frame \OIII\ EWs in 100--300\,\AA\ with a median EW of 194\,\AA, suggestive of the presence of strong \OIII\ emission line. For those with $A_V$ $<1$\,mag, the \OIII\ EW varies from 70\,\AA\ to 400\,\AA.}
\label{f:AV-EW.pdf}
\end{figure}

The \OIII\ EWs of our  sample ELGs vary over a wide range  from 70\,\AA\ to 500\,\AA, with a median value of $\sim$200\,\AA\, and 15\% of them are larger than 300\,\AA.  The relation between $A_V$ derived from CIGALE fitting and EW is shown in Figure~\ref{f:AV-EW.pdf}.   There are eight sample galaxies having $A_V>1\,$mag, showing a relatively high fraction of dusty \OIII\ ELGs.  The majority of the \OIII\ ELGs with $A_V<1\,$mag have \OIII\ EWs over  70$-$400\,\AA.  Further discussion about the rest-frame EWs of our \OIII\ ELGs will be presented in Section~\ref{sec:EELGs}.

%The relation between EW and stellar mass is shown in Figure~\ref{f:EW-M.pdf}.  The increasing \OIII$\lambda5007$ EW$_{\mathrm{rest}}$ with decreasing stellar mass can be explained by stronger star formation activities as well as decreasing metallicity in lower mass SFGs. Further discussion about the rest-frame EWs of our \OIII\ ELGs will be presented in Section~\ref{sec:EELGs}. 

\subsection{Stellar Mass and SFR} \label{sec:MS}

\begin{figure}
\centering
\includegraphics[width=0.45\textwidth]{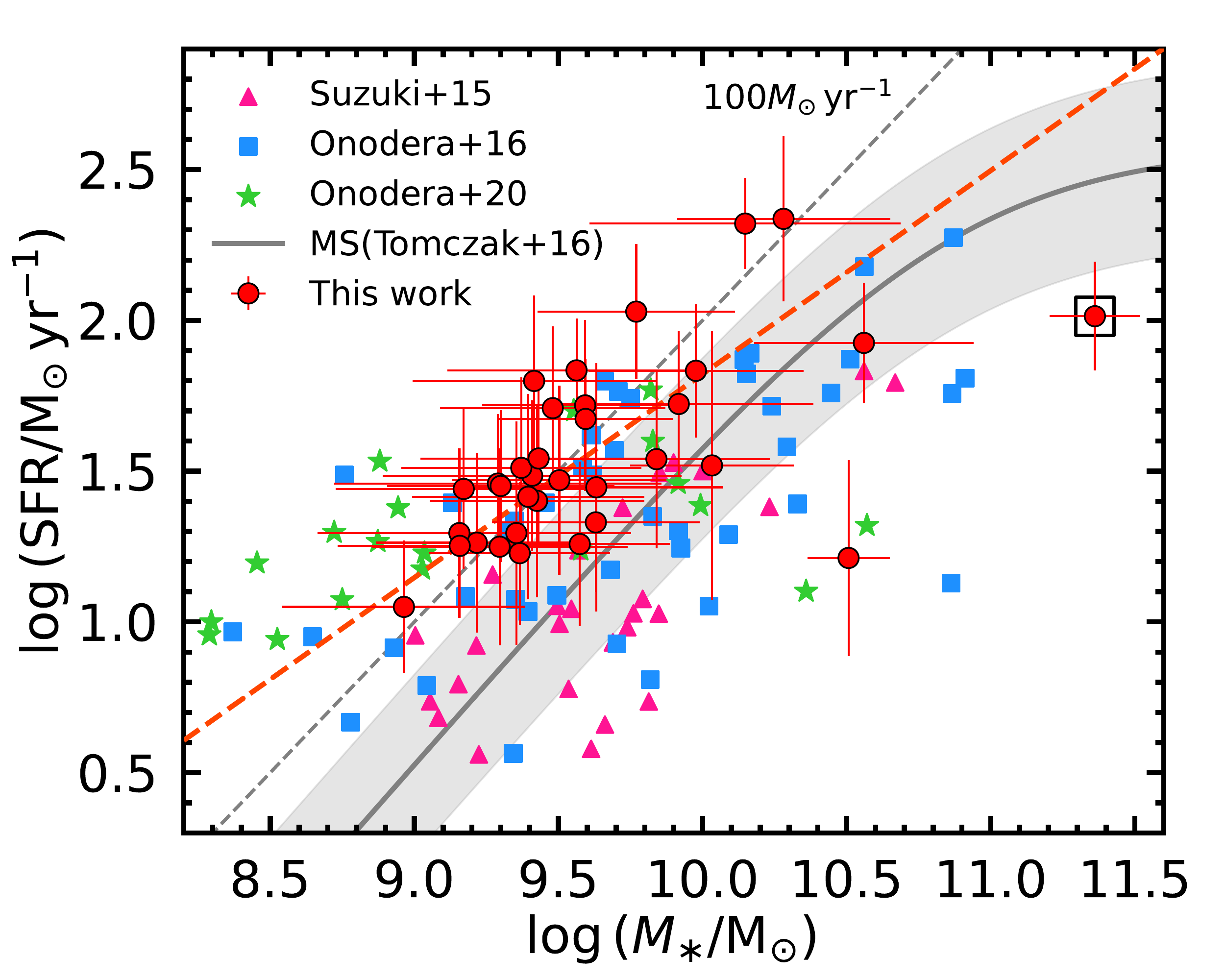}
\caption{Relationship between stellar mass and SFR for our 34 \OIII\ ELGs at $z\sim3.25$ (red circles),
		in comparison with \OIII\ ELGs at $z\sim3.2$ from \citet{Suzuki2015} (pink triangles), 
		at $z\sim3.3$ from \citet{Onodera2016} (blue squares), 
		and at $z\sim3.3$ from \citet{Onodera2020} (green stars).
		The red dashed line is the best fit to our 33 \OIII\ ELGs (excluding the X-ray source), 
		giving a best-fit slope of 0.69. 
		The gray solid curve represents the best fit for the SFMS of SFGs at $z\sim3.25$ given in \citet{Tomczak2016}, 
		with the shaded area showing a 0.3\,dex dispersion. 
		The gray dashed line shows the timescale of 100\,Myr for a galaxy doubling stellar mass. 
		 Most of our sample \OIII\ ELGs lie above the SFMS with higher SFRs.}
\label{f:MS.pdf}
\end{figure}

Galaxy properties estimated through SED fitting with CIGALE are listed in Table~\ref{tab:properties}.  The stellar mass versus SFR relation of our \OIII\ ELG sample is shown in Figure~\ref{f:MS.pdf}.  For a comparison, we also include other samples of \OIII\ ELGs at similar redshifts from the literature, including \OIII\ emitters at $z\sim3.2$ from \citet{Suzuki2015}, and at $z\sim3.3$ from \citet{Onodera2016,Onodera2020}.

We perform an orthogonal distance regression (ODR) fit to the data points of our sample and estimate the dispersion.  For our 33 \OIII\ SFGs, the ODR fit gives a best-fit slope of 0.67 and a median dispersion of 0.15\,dex, while the other three works hold a dispersion of 0.16, 0.23, and 0.16\,dex for \citet{Suzuki2015}, \citet{Onodera2016}, and \citet{Onodera2020}, respectively.  The solid gray curve shows the best-fit relation for the star-forming main sequence (SFMS) of SFGs at $z\sim3.25$ given in \citet{Tomczak2016}, and the shaded area represents a dispersion of 0.3\,dex.  The gray dashed line denotes the timescale of 100\,Myr for a galaxy doubling stellar mass, and the majority of our \OIII\ sample SFGs are located around this line and mostly above the MS.   It is clear that the majority of our \OIII\ ELGs have relatively higher SFRs than typical SFGs of the same masses at similar redshifts.  The dispersion of our sample galaxies is comparable to that of other \OIII\ ELG samples.

In addition, compact SFGs tend to be found at the upper envelope of the SFMS \citep[e.g.,][]{Barro2017,Gomez-Guijarro2019}.  Our \OIII\ ELGs are mainly young and blue SFGs with strong star formation activities, and more than one-third of our sample galaxies are classified as compact ones in terms of their morphologies in the rest-frame UV.   We thus argue that the compact \OIII\ ELGs in our sample resemble compact SFGs at $z\sim3.25$ as young starburst galaxies with high SFRs.

\subsection{Dust Attenuation} \label{sec:attenuaion}

The empirical dust attenuation curve of \citet{Calzetti2000} is widely used for starburst galaxies.  \citet{Noll2009b} modified the Calzetti law by multiplying a power-law function with a slope of $\delta$ and adding a  2175\AA\ bump that is described by a Lorentzian-like Drude profile.  We take the modified Calzetti law as the dust attenuation curve in our analysis and derive dust attenuation $A_V$ from the SED fitting with CIGALE.  In practice, the stellar color excess $E(B-V)_\mathrm{star}$ and attenuation curve slope $\delta$ are set as free parameters in the fitting.

From the best-fit results with CIGALE, we obtain $E(B-V)_\mathrm{star}$, $\delta$, and the strength of the  2175\AA\ bump for each sample galaxy.  We are able to estimate the attenuation at a given band using the global attenuation formula from \citet{Boquien2019} as 
\begin{equation} \label{e:modatt}
	k_{\lambda} = (k_\lambda^{starburst}\times(\lambda/550~\textrm{nm})^\delta+D_\lambda)\times\frac{\mathrm{E(B-V)_{\delta=0}}}{\mathrm{E(B-V)_{\delta}}},
\end{equation}
where $D_\lambda$ is the Drude profile that describes the 2175\AA\ bump, and the last term renormalizes the curve so that $E(B-V)$ remains equal to the input $E(B-V)$ when $\delta\neq 0$.  We then estimate the attenuation to the \OIII$\lambda5007$ line $A_\mathrm{\OIII}$ with this equation, and the results are shown in Table~\ref{tab:properties}.

It can be clearly seen from  Figure~\ref{f:sed.pdf} that the 2175\AA\ bump is present in the best-fit SEDs of seven sample galaxies (e.g., ID=5 and 13). The 2175\AA\ bump is redshifted to around 9250\,\AA\ at $z\sim3.25$, corresponding to $I$ and $z_{850}$. These seven galaxies have relatively high SFRs and $A_\mathrm{\OIII}$. Moreover, half of them are compact, and the other half are extended in morphology.  We do not see a connection between the presence of the 2175\AA\ bump and galaxy morphology among these $z\sim 3.25$ \OIII\ ELGs.

Previous studies reported that the 2175\AA\ bump is commonly seen in SFGs up to $z\sim2.6$ \citep{Buat2011,Wild2011,Shivaei2020,Kashino2021}.  \citet{Noll2009a} pointed out that at least 30\% of SFGs at $1<z<2.5$ exhibit a significant 2175\AA\ bump.  In our SED fitting, the 2175\AA\ bump is introduced when the $z_{850}$ flux is lower than the $I$ flux in an SED.  We caution that either an overestimate of the $I$ flux or an underestimate of the $z_{850}$ flux might demand a stronger  2175\AA\ bump in the models for a better fit.  We examine the 2175\AA\ bump of those best-fit SEDs, finding that most of the seven sample galaxies have a higher $I$ flux than the $z_{850}$.  Due to the lack of spectroscopic and more photometric data, it is difficult to securely confirm the 2175\AA\ bump. More efforts are necessary to investigate the origin of the 2175\AA\ bump in $z\sim 3.25$ \OIII\ ELGs.

\begin{figure}
\centering
\includegraphics[width=0.45\textwidth]{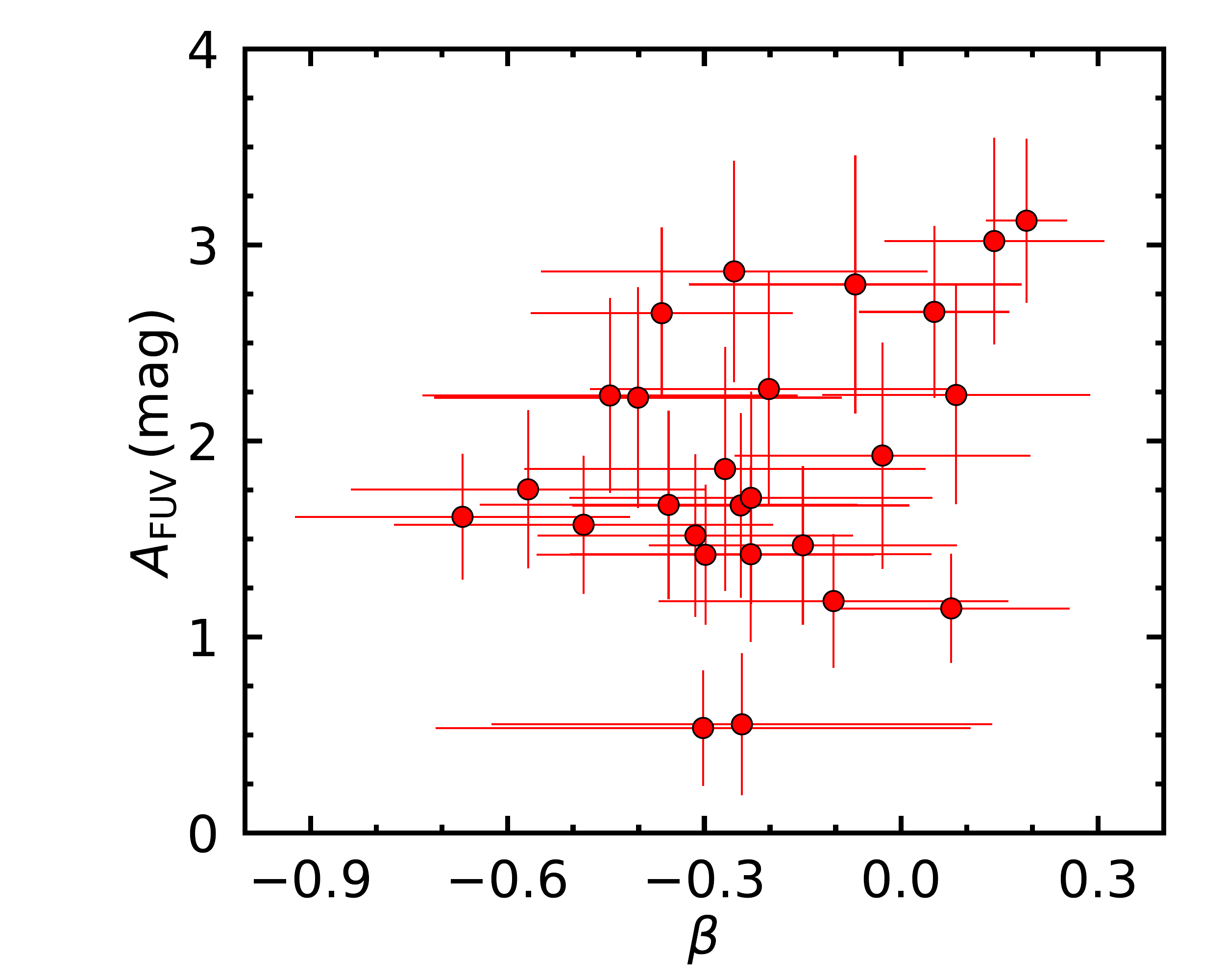}
\caption{UV slope $\beta$ versus dust attenuation in the FUV $A_\mathrm{FUV}$ for our sample of 26 \OIII\ ELGs.  
  Eight faint sources with upper limits in both $U$ and $B$ are ignored since the measurements of their $A_\mathrm{FUV}$ and $\beta$ are largely uncertain.
  The scatter is large owing to the lack of UV photometric data.}
\label{f:AFUV-beta.pdf}
\end{figure}

We estimate $A_\mathrm{FUV}$ at 1500\,\AA\ with Equation~\ref{e:modatt} and show the correlation between $A_\mathrm{FUV}$ and UV slope $\beta$ in Figure~\ref{f:AFUV-beta.pdf}.  There are eight sample galaxies that are compact ($R<0.1\arcsec$), faint, or nearly invisible in the rest-frame UV (see Figure~\ref{f:ACS.pdf}), indicating a highly obscured environment around these galaxies.  In addition, these sources also show upper limits in both $U$ and $B$ in their SEDs (see Figure~\ref{f:sed.pdf}); therefore, the UV slope $\beta$ becomes meaningless for such a case.  Thus, these eight sources are not included in Figure~\ref{f:AFUV-beta.pdf}.  The UV slope $\beta$ is often used as a measure of dust obscuration in the sense that a redder UV slope (a higher $\beta$) is linked to a higher dust attenuation \citep{Meurer1999}.  From Figure~\ref{f:AFUV-beta.pdf}, however, we show that our sample galaxies do not exhibit a clear correlation between $\beta$ and $A_{\rm FUV}$.  As can be seen, some of our sample galaxies are barely visible in the rest-frame FUV owing to the heavy dust attenuation, and the estimates of $\beta$ and $A_{\rm FUV}$ could be significantly biased by the leaking UV radiation.

\subsection{\OIII\ Luminosity} \label{sec:luminosity}

Following \citet{Ly2011}, we estimate the line flux of \OIII$\lambda$5007 from the narrowband excess using 
\begin{equation} \label{flux}
	F_\mathrm{[OIII]} = \Delta{\mathrm{NB}}\,\frac{f_\mathrm{NB} - f_\mathrm{K_\mathrm{s}}} {1-(\Delta{\mathrm{NB}}/\Delta K_\mathrm{s})},
\end{equation}
where $F_\mathrm{[OIII]}$ is the integrated line flux of \OIII\ given in units of erg\,s$^{-1}$\,cm$^{-2}$, and $f_\mathrm{NB}$ and $f_{K_\mathrm{s}}$ refer to fluxes in H$_2\mathrm{S}(1)$ and $K_\mathrm{s}$, respectively, given in the units of erg\,s$^{-1}$\,cm$^{-2}$\,\AA$^{-1}$.  The band widths are $\Delta\mathrm{NB}=293$\,\AA\ and $\Delta K_\mathrm{s}=3250$\,\AA.   We note that our sample was selected with EW$>50$\,\AA\ and the 5$\sigma$ detection limits are H$_2\mathrm{S}(1)=22.8$\,mag and $K_\mathrm{s}=24.8$\,mag.  The detection limit for the \OIII\ emission line is then estimated to be $>2.5\times 10^{-17}$\,erg\,s$^{-1}$\,cm$^{-2}$.  These luminosities are corrected for dust attenuation with $A_\mathrm{\OIII}$.  We calculate the \OIII\ line luminosity for our sample galaxies and present them in Table~\ref{tab:properties}.

\begin{figure}
\centering
\includegraphics[width=0.45\textwidth]{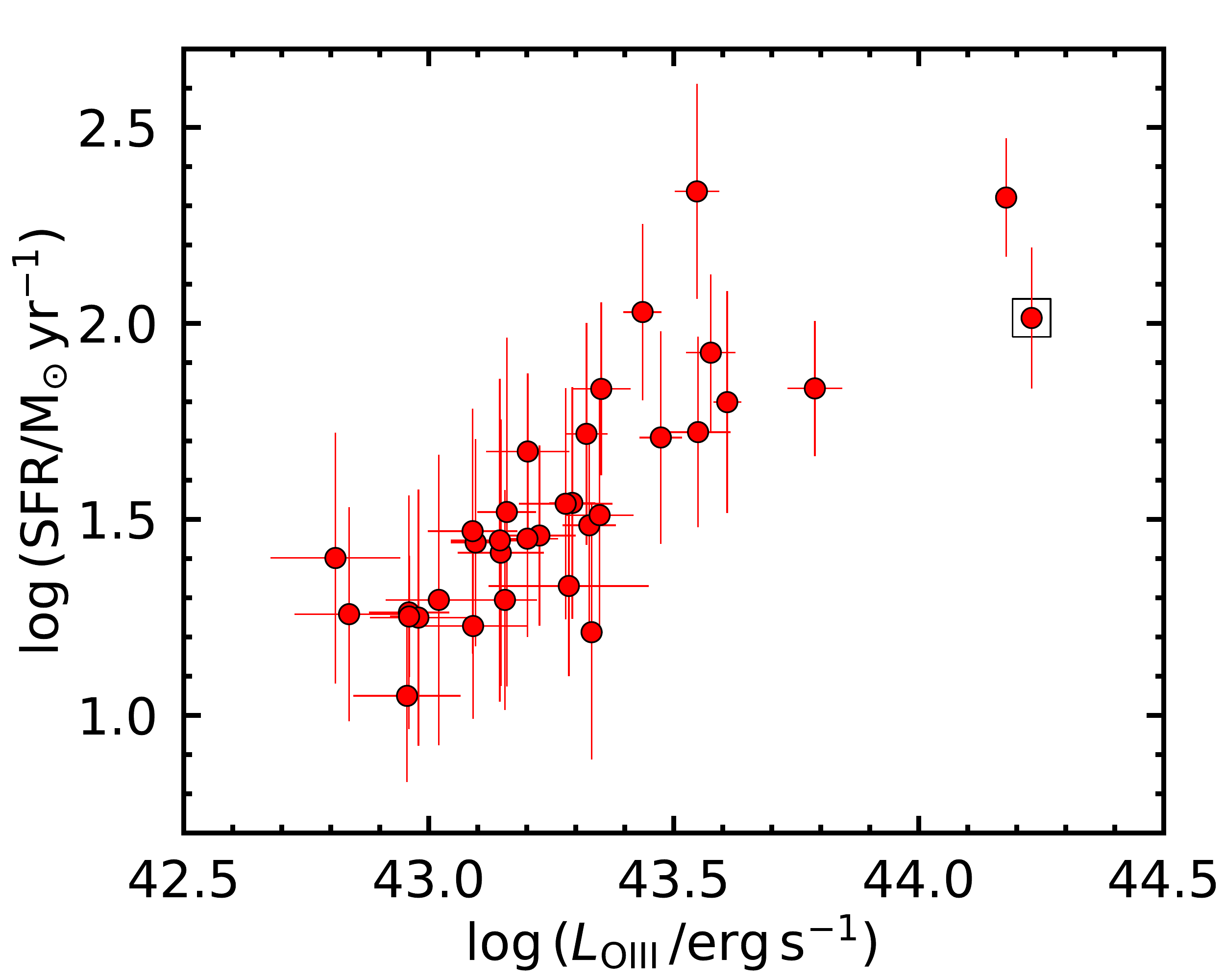}
\caption{Relationship between \OIII\ luminosity and SFR.
		The \OIII\ luminosities are corrected for dust attenuation obtained from CIGALE. 
		It is clear that SFR increases with \OIII\ luminosity in a statistic manner.}
\label{f:SFR-L.pdf}
\end{figure}

Figure~\ref{f:SFR-L.pdf} shows the relation between SFR and \OIII\ luminosity.  A strong correlation is seen. Such a correlation has been reported before \citep[e.g.][]{Straughn2009,Villa-Velez2021}.  It is also clear that most of our \OIII\ ELGs have dust-corrected luminosities in the range of $10^{42.6}-10^{43.6}$.  In order to show the representativeness of luminosity for our \OIII\ ELGs, the luminosity functions of \OIII\ ELGs at $z\sim3.24$ from \citet{Khostovan2015} are adopted for comparison.  Note that the luminosity function parameter results in their work are uncorrected for dust and AGN contribution owing to the undeveloped roles of dust on emission lines at the high-$z$ universe.  The characteristic luminosity of \OIII\ ELGs at $z\sim3.24$ in \citet{Khostovan2015} is $\log\,L_\ast=42.83$\,erg\,s$^{-1}$, while the median value of the uncorrected luminosity of our \OIII\ sample is $10^{42.73}$\,erg\,s$^{-1}$, showing a variance of 0.1\,dex.  The similarity in uncorrected \OIII\ luminosity indicates that our \OIII\ sample shows a typical property of luminosity at this redshift.

\begin{deluxetable*}{lccccccccccccc}
\tablecaption{Physical Properties of 34 Sample \OIII\ Emitters \label{tab:properties}}
\tablewidth{0pt}
\tabletypesize{\scriptsize}
\centering 
%\scalebox{0.72}{
\tablehead{
\colhead{ID} & \colhead{$z_\mathrm{phot}$} & \colhead{$z_\mathrm{spec}$} & \colhead{$T^\mathrm{a}$} & \colhead{$U-V^\mathrm{b}$} & \colhead{$V-J^\mathrm{c}$} & \colhead{EW$_\mathrm{\OIII rest}$} & \colhead{$\log\,M_{\ast}$} & \colhead{$\log\,\mathrm{SFR}$} & \colhead{$\beta$} & \colhead{$A_\mathrm{FUV}$} & \colhead{$A_\mathrm{V}$}  &\colhead{$\log\,L_\mathrm{\OIII}$} \\
 & & & & \colhead{(mag)} & \colhead{(mag)} & \colhead{(\AA)} & \colhead{(M$_\odot$)} &  \colhead{(M$_\odot$\,yr$^{-1}$)} & & \colhead{(mag)} & \colhead{(mag)} & \colhead{(erg\,s$^{-1}$)}}
%\decimalcolnumbers
\startdata
1   & 3.260 &  ...  &  3  & 0.36 &   0.32  &  94.64$\pm$36.60 &  9.43$\pm$0.37 & 1.40$\pm$0.32 & $-$0.48$\pm$0.29 & 1.57$\pm$0.35 & 0.49$\pm$0.27 & 42.77$\pm$0.13 \\
2   & 3.256 & 3.083 &  3  & 0.70 &   0.54  & 133.52$\pm$25.26 & 10.03$\pm$0.28 & 1.52$\pm$0.44 & $-$0.40$\pm$0.31 & 2.22$\pm$0.56 & 0.67$\pm$0.43 & 43.09$\pm$0.06 \\
3   & 3.254 &  ...  &  4  & 0.46 &   0.07  & 233.71$\pm$41.70 &  9.48$\pm$0,39 & 1.71$\pm$0.27 & $-$0.24$\pm$0.26 & 1.67$\pm$0.47 & 0.74$\pm$0.38 & 43.45$\pm$0.04 \\
4   & 3.260 &  ...  &  5  & 0.51 &   0.21  & 141.08$\pm$49.01 &  9.35$\pm$0.40 & 1.29$\pm$0.37 & $-$0.35$\pm$0.29 & 1.67$\pm$0.48 & 0.63$\pm$0.38 & 42.98$\pm$0.11 \\
5   & 3.254 &  ...  &  3  & 0.62 &   0.79  &  79.96$\pm$10.98 & 10.28$\pm$0.37 & 2.34$\pm$0.27 &    0.05$\pm$0.11 & 2.66$\pm$0.44 & 1.30$\pm$0.35 & 43.55$\pm$0.05 \\
6   & 3.254 &  ...  &  3  & 0.57 &   0.20  & 362.89$\pm$58.42 &  9.42$\pm$0.42 & 1.80$\pm$0.28 & $-$0.20$\pm$0.27 & 2.27$\pm$0.59 & 1.08$\pm$0.48 & 43.60$\pm$0.03 \\
7   & 3.256 &  ...  &  2  & 0.50 & $-$0.01 & 291.15$\pm$71.03 &  9.17$\pm$0.44 & 1.44$\pm$0.26 &    ...  & ... & 0.68$\pm$0.35 & 43.10$\pm$0.05 \\
8   & 3.254 &  ...  &  3  & 0.53 &   0.34  & 138.95$\pm$18.62 &  9.77$\pm$0.34 & 2.03$\pm$0.22 & $-$0.36$\pm$0.20 & 2.65$\pm$0.44 & 0.95$\pm$0.34 & 43.39$\pm$0.04 \\
9   & 3.256 &  ...  &  3  & 0.42 &   0.03  & 196.07$\pm$56.84 &  9.22$\pm$0.35 & 1.26$\pm$0.30 & $-$0.67$\pm$0.26 & 1.61$\pm$0.32 & 0.45$\pm$0.25 & 42.93$\pm$0.08 \\
10$^{\rm d}$ & 3.256 &  ...  & 2 & 1.61 &  0.93 & 213.02$\pm$5.69 & 11.36$\pm$0.16 & 2.01$\pm$0.18 & ... & ... & 0.79$\pm$0.16 & 44.24$\pm$0.01 \\
11  & 3.256 &  ...  &  3  & 0.65 &   0.36  & 255.78$\pm$78.04 &  9.29$\pm$0.57 & 1.46$\pm$0.23 &    0.08$\pm$0.20 & 2.23$\pm$0.56 & 1.30$\pm$0.45 & 43.22$\pm$0.07 \\
12  & 3.253 &  ...  &  2  & 0.66 &   0.19  & 278.28$\pm$63.18 &  9.41$\pm$0.52 & 1.48$\pm$0.25 &    ...  & ...  & 0.89$\pm$0.38 & 43.32$\pm$0.05 \\
13  & 3.254 & 3.232 &  2  & 0.85 &   0.71  & 483.14$\pm$44.47 & 10.15$\pm$0.54 & 2.32$\pm$0.15 &    0.19$\pm$0.06 & 3.12$\pm$0.42 & 1.87$\pm$0.35 & 44.17$\pm$0.01 \\
14  & 3.253 & 3.244 &  5  & 0.49 &   0.23  & 172.30$\pm$53.99 &  9.40$\pm$0.40 & 1.42$\pm$0.34 & $-$0.23$\pm$0.28 & 1.71$\pm$0.54 & 0.74$\pm$0.43 & 43.10$\pm$0.09 \\
15  & 3.253 & 3.248 &  4  & 0.33 &   0.26  & 159.33$\pm$39.93 &  9.37$\pm$0.42 & 1.51$\pm$0.30 & $-$0.30$\pm$0.26 & 1.42$\pm$0.36 & 0.57$\pm$0.28 & 43.31$\pm$0.07 \\
16  & 3.256 &  ...  &  3  & 0.55 &   0.14  & 180.40$\pm$63.89 &  9.30$\pm$0.44 & 1.25$\pm$0.33 &    ...  & ...  & 0.64$\pm$0.39 & 42.96$\pm$0.10 \\
17  & 3.262 &  ...  &  3  & 0.59 &   0.28  & 129.54$\pm$38.17 &  9.50$\pm$0.37 & 1.47$\pm$0.31 & $-$0.44$\pm$0.29 & 2.23$\pm$0.50 & 0.76$\pm$0.39 & 43.08$\pm$0.09 \\
18  & 3.254 & 3.225 &  4  & 0.48 & $-$0.01 & 317.15$\pm$96.49 &  9.16$\pm$0.49 & 1.29$\pm$0.28 & $-$0.23$\pm$0.28 & 1.42$\pm$0.45 & 0.66$\pm$0.36 & 43.09$\pm$0.07 \\
19  & 3.254 &  ...  &  2  & 0.44 &   0.25  & 160.80$\pm$63.35 &  9.37$\pm$0.31 & 1.23$\pm$0.24 &    0.08$\pm$0.18 & 1.15$\pm$0.28 & 0.60$\pm$0.22 & 43.06$\pm$0.11 \\
20  & 3.254 &  ...  &  1  & 0.89 &   0.47  &  97.56$\pm$46.58 &  9.63$\pm$0.36 & 1.33$\pm$0.23 &    ...  & ...  & 0.87$\pm$0.27 & 43.30$\pm$0.16 \\	
21  & 3.254 &  ...  &  2  & 0.82 &   0.50  &  98.19$\pm$17.73 &  9.98$\pm$0.37 & 1.83$\pm$0.22 &    ...  & ...  & 1.04$\pm$0.34 & 43.37$\pm$0.06 \\
22  & 3.243 & 3.208 &  3  & 0.38 &   0.04  & 112.46$\pm$38.68 &  9.57$\pm$0.31 & 1.26$\pm$0.27 & $-$0.30$\pm$0.41 & 0.54$\pm$0.30 & 0.23$\pm$0.23 & 42.85$\pm$0.11 \\
23  & 3.254 & 3.229 &  5  & 0.46 &   0.04  & 158.91$\pm$22.32 &  9.59$\pm$0.36 & 1.72$\pm$0.28 & $-$0.31$\pm$0.24 & 1.52$\pm$0.42 & 0.62$\pm$0.33 & 43.28$\pm$0.04 \\
24  & 3.256 &  ...  &  2  & 0.50 &   0.00  & 228.56$\pm$40.00 &  9.43$\pm$0.41 & 1.54$\pm$0.30 & $-$0.57$\pm$0.27 & 1.75$\pm$0.40 & 0.59$\pm$0.32 & 43.28$\pm$0.05 \\
25  & 3.251 &  ...  &  3  & 0.76 &   0.56  & 127.21$\pm$37.46 &  9.84$\pm$0.39 & 1.54$\pm$0.30 & $-$0.25$\pm$0.29 & 2.87$\pm$0.56 & 1.10$\pm$0.46 & 43.24$\pm$0.10 \\
26  & 3.253 &  ...  &  2  & 0.46 &   0.06  & 235.85$\pm$63.68 &  9.30$\pm$0.39 & 1.45$\pm$0.25 & $-$0.15$\pm$0.23 & 1.47$\pm$0.40 & 0.70$\pm$0.32 & 43.17$\pm$0.06 \\
27  & 3.254 &  ...  &  2  & 0.83 &   0.62  & 162.22$\pm$37.54 &  9.92$\pm$0.47 & 1.72$\pm$0.24 & $-$0.07$\pm$0.25 & 2.80$\pm$0.66 & 1.37$\pm$0.54 & 43.54$\pm$0.07 \\
28  & 3.250 & 3.217 &  2  & 1.00 &   0.51  & 264.72$\pm$20.56 & 10.51$\pm$0.14 & 1.21$\pm$0.32 & $-$0.24$\pm$0.38 & 0.55$\pm$0.36 & 0.24$\pm$0.27 & 43.34$\pm$0.02 \\
29  & 3.252 &  ...  &  2  & 0.64 &   0.28  & 108.63$\pm$27.47 &  9.59$\pm$0.30 & 1.67$\pm$0.20 &    ...  & ...  & 0.78$\pm$0.23 & 43.17$\pm$0.08 \\
30  & 3.251 &  ...  &  3  & 0.43 &   0.00  & 357.10$\pm$194.67 &  8.96$\pm$0.42 & 1.05$\pm$0.22 & $-$0.10$\pm$0.27 & 1.18$\pm$0.34 & 0.59$\pm$0.27 & 42.89$\pm$0.11 \\
31  & 3.260 &  ...  &  3  & 0.62 &   0.29  & 121.42$\pm$37.10 &  9.63$\pm$0.44 & 1.45$\pm$0.41 & $-$0.27$\pm$0.31 & 1.86$\pm$0.62 & 0.78$\pm$0.50 & 43.11$\pm$0.10 \\
32  & 3.260 &  ...  &  2  & 0.78 &   0.65  & 336.53$\pm$92.54 &  9.56$\pm$0.45 & 1.83$\pm$0.17 &    0.14$\pm$0.17 & 3.02$\pm$0.53 & 1.79$\pm$0.43 & 43.79$\pm$0.06 \\
33  & 3.254 &  ...  &  1  & 1.09 &   0.56  &  74.43$\pm$11.17 & 10.56$\pm$0.38 & 1.93$\pm$0.20 &    ...  & ...  & 0.83$\pm$0.22 & 43.58$\pm$0.05 \\
34  & 3.251 &  ...  &  5  & 0.31 &   0.47  &  91.94$\pm$18.67 &  9.16$\pm$0.42 & 1.25$\pm$0.15 & $-$0.03$\pm$0.23 & 1.93$\pm$0.58 & 0.96$\pm$0.47 & 42.67$\pm$0.04 \\
\enddata
%\vskip 0.03in 
\noindent {\scriptsize $^{a}$ Morphology Type: 1--UV faint; 2--compact; 3--diffuse/clumpy/tidal; 4--merger; 5--multiple components.} \\
\noindent {\scriptsize $^{b}$ A typical error of $U-V$ is 0.12.} \\
\noindent {\scriptsize $^{c}$ A typical error of $V-J$ is 0.24.} \\
\noindent {\scriptsize $^{d}$ X-ray source (with XID=760 in the \textit{Chandra} 7\,Ms catalog).} \\

\end{deluxetable*}

\section{An Overdensity Traced by \OIII\ ELGs} \label{sec:overdensity}

\begin{figure*}
\includegraphics[width=0.9\textwidth]{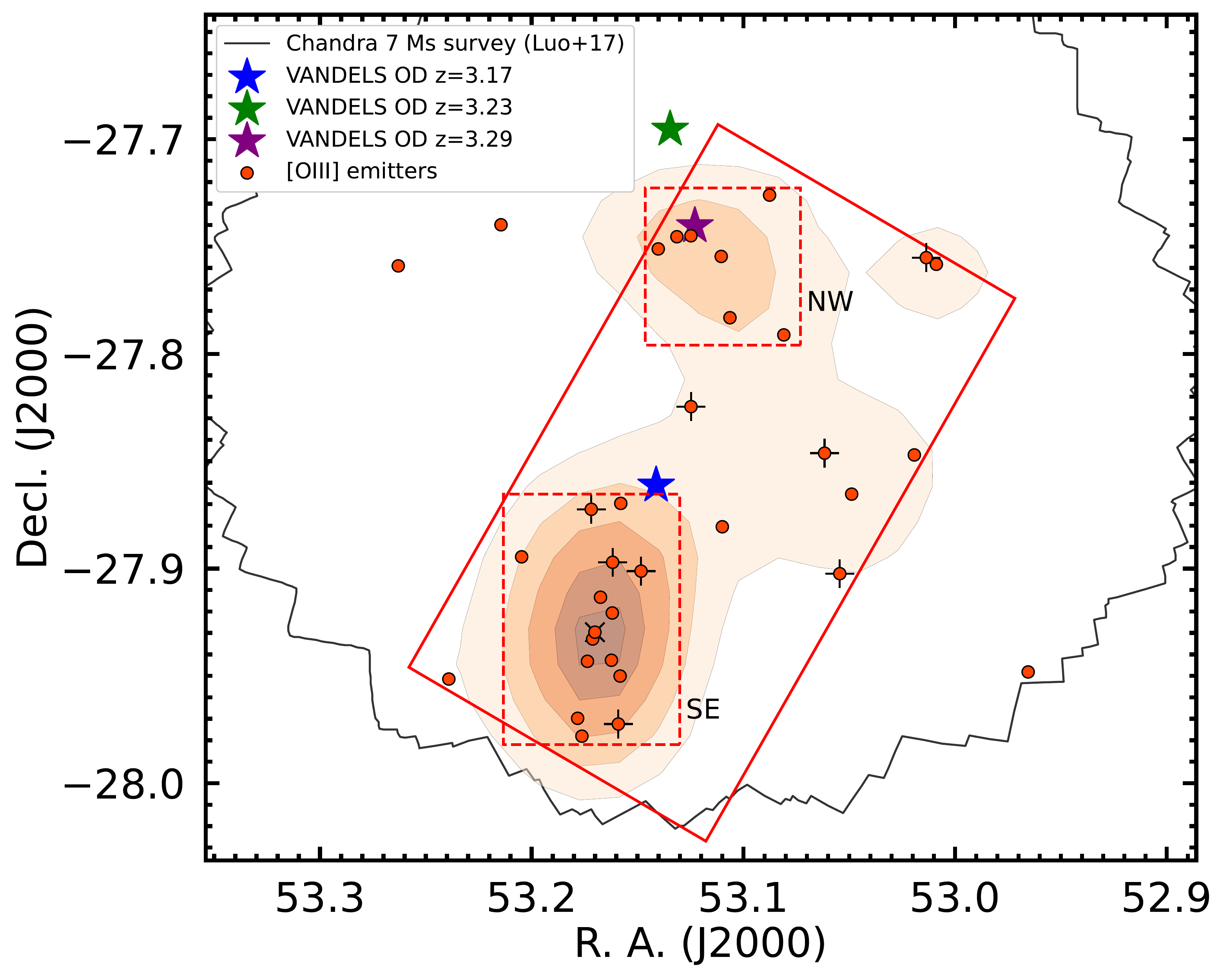}
\caption{Spatial distribution of 34 \OIII\ ELGs at $z\sim3.25$ in ECDFS. 
    Red circles represent our \OIII\ ELGs and pluses denote the spectroscopically-confirmed ones.  
    The cross symbol  marks the AGN source. 
    The contour levels of the density map refer to [4, 8, 12, 16, 20] $\times$ the \OIII\ emitter surface number density of general fields ($2.7\times10^{-2}$\,arcmin$^{-2}$). 
    The red solid rectangle covers 31 sources with an area of $17\farcm5 \times 9\farcm7$. 
    The South-East (SE) component ($7\arcmin \times 5\arcmin$) and the North-West (NW) component ($4\farcm4 \times 4\farcm4$) covers 15 and 7 sources, respectively. 
    The stars mark the locations of identified overdensities traced by Lyman-$\alpha$ emitters at similar redshifts in CDFS from the VANDELS survey \citet{Guaita2020}.
    Black solid lines show the coverage of \textit{Chandra} 7\,Ms X-ray observations.}
\label{f:SD.pdf}
\end{figure*}

Figure~\ref{f:SD.pdf} shows the spatial distribution of our sample of 34 \OIII\ ELGs. It is obvious that they are strongly clustered and form an overdensity.  We estimate its overdensity factor as $\delta_\mathrm{gal} = N_\mathrm{group}/N_\mathrm{field} - 1$.  The effective detection area of our sample is 383\,arcmin$^2$. The narrowband filter H$_2\mathrm{S}(1)$ ($\lambda_\mathrm{c}=2.130\,\micron$, $\Delta\lambda=0.0293\,\micron$) covers a redshift span of $z=3.254\pm0.029$ for \OIII$\lambda$5007, corresponding to a radial scale of 50.9\,comoving\,Mpc (cMpc). 
The total comoving volume of ECDFS is then estimated to be 41.5$^3$\,cMpc$^3$.

\subsection{General Field Number Density} \label{sec:generalfield}

In order to estimate the overdensity factor of \OIII\ ELGs in ECDFS, we firstly estimate the number density in general fields.  The High-redshift(Z) Emission Line Survey (HiZELS) provides a large sample of ELGs identified by narrowband excesses in COSMOS and UDS \citep{Sobral2013}.  The NB$_K$ filter ($\lambda_\mathrm{c}=2.1210\,\micron$, $\Delta\lambda=0.0210\,\micron$) probes \OIII\ ELGs at $z\sim3.24$, which is similar to our sample galaxies.  Given that the improved photometric catalog is available in COSMOS, we cross the NB$_K$-excess object catalog from \citet{Sobral2013} with the COSMOS2020 catalog \citep{Weaver2021} to identify the \OIII\ ELGs in COSMOS.  Following the selection criteria used in \citet{Khostovan2015}, we pick in total 159 NB$_K$-excess objects with $2.8\leq z_\mathrm{phot}\leq 4$ to be \OIII\ ELGs.  We use this sample to estimate the number density of \OIII\ ELGs at $z\sim 3.24$ in general fields.

The selection criterion for ELGs (EW$>$25\,\AA) from \citet{Sobral2013} differs from ours (EW$>$50\,\AA).  And the detection limits on narrow- and broadband depths also affect the selection for \OIII\ emitters.  We apply our \OIII\ ELG selection criteria (EW$>$50\,\AA, H$_2$S(1)$<22.8$\,mag, and $K_{\rm s}<24.8$\,mag) to the selected 159 \OIII\ emitters in COSMOS, given that 157 of the 159 \OIII\ emitters meet our selection criteria.  \OIII\ ELGs at this redshift usually have a high EW; therefore, the difference between two EW cuts has a negligible impact on the selection. The effective sky coverage for these \OIII\ ELGs is $\sim1.6\,$deg$^{2}$, while the NB$_K$ filter covers a comoving distance in the line of sight with 37.1\,cMpc.  So the total volume of these 157 \OIII\ emitters is about $7.8\times10^{5}$\,cMpc$^3$.  We estimate the number density per volume of \OIII\ ELGs in COSMOS to be $2.0\times10^{-4}$\,cMpc$^{-3}$. \citet{Khostovan2015} showed an \Hb+ \OIII\ number density of $2.3\times10^{-4}$\,cMpc$^{-3}$ at $2.8\leq z_\mathrm{phot}\leq 4$ in COSMOS and UDS, while \citet{Forrest2017} presented the strong \OIII\ ELG number density of $2.3\times10^{-4}$\,cMpc$^{-3}$ at $2.5\leq z_\mathrm{phot}\leq 4$ in CDFS.  This indicates that our number densities are comparable to other studies.  So we take this to be the number density of $z=3.25$ \OIII\ ELGs in general fields.

\subsection{Density Map and $\delta_\mathrm{gal}$} \label{sec:densitymap}

Our sample includes 34 \OIII\ ELGs detected over 383\,arcmin$^2$ in ECDFS and redshift span of $z=3.254\pm0.029$, giving a number density of \OIII\ ELGs $(4.8\pm0.8)\times10^{-4}$\,cMpc$^{-3}$.  We note that our sample \OIII\ ELGs are located in a smaller area than the detection area. We construct a density map for our $z\sim3.25$ \OIII\ ELGs and estimate the coverage area of the overdensity in ECDFS.  Following \citet{Zheng2021}, the detection area is divided into a grid of $1\arcmin \times 1\arcmin$ cells, and the number of ELGs is counted in each cell to obtain the number density.  A Gaussian kernel of $\sigma=1\arcmin$ (i.e., 1.9\,cMpc at $z\sim3.25$) is used to convolve the grid and yield the density map as shown in Figure~\ref{f:SD.pdf}.  The contour levels are drawn at the 4, 8, 12, 16, and 20 $\times$ the surface \OIII\ emitter number density of general fields.

Clearly, the majority of our sample galaxies are located in the central region of ECDFS. The red solid rectangle in Figure~\ref{f:SD.pdf} encloses 31 \OIII\ ELGs  over an area of 170.7\,arcmin$^2$ ($17\farcm5 \times 9\farcm7$).  The density map of \OIII\ ELGs consists of two components: one southeast (SE) composed of 15 \OIII\ ELGs over 35.0\,arcmin$^2$ ($7\arcmin \times 5\arcmin$), and the other northwest (NW) with seven sample galaxies over 19.4\,arcmin$^2$ ($4\farcm4 \times 4\farcm4$).  Four spectroscopically identified \OIII\ ELGs are also located in the SE component, confirming this overdensity region is located at $z\sim3.25$.

We then estimate the volume number density of \OIII\ ELGs in these components. We note that the redshift span of $z=3.254\pm0.029$ used to calculate the radial scale may be too large for calculating the comoving volumes because our sample \OIII\ ELGs unlikely fulfill the band width of H$_2$S(1) and the actual radial size of the \OIII\ overdensity is smaller than 50.9\,cMpc.  Given that the available spec-$z$ of our sample \OIII\ ELGs are all at $z<3.255$, the half of the redshift span covered by the H$_2$S(1) filter (see Figure~\ref{f:zdistribution.pdf}), we thus take half of the radial comoving distance (25.5\,cMpc) to be the upper limit for the radial size of the \OIII\ overdensity.  On the other hand, the radial size of the overdensity is unlikely smaller than the the minor axis of each component area, giving a lower limit of the radial comoving distance.  Therefore, we take the the lower and upper limits of the radial size to estimate the comoving volume for these three components.

The red solid rectangle area covers a comoving volume from 25$^3$\,cMpc$^3$ to $23^3$\,cMpc$^3$.  We then estimate the \OIII\ overdensity to have a volume number density from $(1.9\pm0.4)\times10^{-3}$\,cMpc$^{-3}$ to $(2.7\pm0.5)\times10^{-3}$\,cMpc$^{-3}$, giving the $\delta_\mathrm{gal}$ in the range of [$8.7\pm2.0$, $12.3\pm2.8$].  The SE component contains 15 objects in a comoving volume from $15^3$\,cMpc$^3$ to $11^3$\,cMpc$^3$, giving a volume number density from $(4.6\pm1.2)\times10^{-3}$\,cMpc$^{-3}$ to $(12.1\pm3.2)\times10^{-3}$\,cMpc$^{-3}$ and $\delta_\mathrm{gal}$ in the range of $[21.8\pm5.7, 59.5\pm16.9]$.  The NW component contains seven objects in the volume from $12^3$\,cMpc$^3$ to $8^3$\,cMpc$^3$, yielding a number density of $(3.9\pm1.7)\times10^{-3}$\,cMpc$^{-3}$ to $(11.7\pm5.0)\times10^{-3}$\,cMpc$^{-3}$.  And the $\delta_\mathrm{gal}$ of NW 
varies in the range of [$18.2\pm8.4$, $57.1\pm25.4$].  The uncertainties are simply estimated by shot noise.  These estimated overdensity factors confirm that our sample \OIII\ ELGs in ECDFS reside in an overdensity.

\subsection{Present-day Mass}\label{sec:Matz0}

The typical size of a protocluster at $z=3$ is about 20\,cMpc \citep{Chiang2013}.  We estimate the expected total mass at $z\sim0$ for the overdensity components at $z=3.25$ in ECDFS using  
\begin{equation}
   M_{z=0} = \bar{\rho}\,(1 + \delta_\mathrm{m})\,V_\mathrm{true},
\end{equation}
from \citet{Steidel1998}.  Here $\bar{\rho}$ is the mean comoving matter density of the universe, which equals to $\frac{3 \mathrm{H_0^{2}}}{\mathrm{8\pi G}} = 4.1\times10^{10}\,\mathrm{M_{\sun}}\,\mathrm{cMpc^{-3}}$,  $\delta_\mathrm{m}$ is the matter overdensity; and $V_\mathrm{true} = V_\mathrm{obs}/C$. From \citet{Steidel1998}, $V_\mathrm{obs}$ is the observed comoving volume and $C$ is a correction factor estimated using $C = 1 + f - f (1+\delta _\mathrm{m})^{1/3}$, where $f = \Omega_\mathrm{m} z^{4/7}$, and $f=0.98$ at $z=3.25$.  And $\delta_\mathrm{m}$ is linked to the galaxy overdensity by $1 + b\,\delta_\mathrm{m} = C\,(1+\delta _\mathrm{gal})$, where $b$ is the \OIII\ emitter bias factor.  We adopt the linear bias $b=3.43$ for \OIII\ ELGs in the redshift range of 2--3 from \citet{Zhai2021} as the bias for \OIII\ ELGs at $z=3.25$.

We calculate the correction factor $C$ and matter overdensity $\delta_\mathrm{m}$ for two overdensity components. For the SE substructure, we obtain $\delta_\mathrm{m}=4.33$ and $C=0.27$ for the lower limits and $\delta_\mathrm{m}=2.67$ and $C=0.47$ for the upper limits.  For the NW substructure, we obtain $\delta_\mathrm{m}=4.26$ and $C=0.27$ for the lower limits and $\delta_\mathrm{m}=2.39$ and $C=0.51$ for the upper limits.  And for the entire structure, we get $\delta_\mathrm{m}=1.83, 1.40$ and $C=0.59, 0.67$ for the lower and upper limits, respectively.  So the present-day mass is then estimated to be $\sim1.1\times10^{15}\,$\,M$_\odot$ for the SE substructure and $\sim4.8\times10^{14}\,$\,M$_\odot$ for the NW substructure.  And the present-day mass of the entire structure is $\sim2.3\times10^{15}$\,M$_\odot$.

Based on these estimates, we conclude that the overdensity traced by our \OIII\ ELGs is indeed a massive protocluster of galaxies at $z\sim 3.25$ in ECDFS.  These two substructures are expected to become virialized at $z=0$, with the SE substructure probably being a high-mass ``Coma-type'' cluster of $\sim10^{15}$\,M$_\odot$ and the NW substructure forming an intermediate-mass ``Virgo-type'' cluster of (3--9)\,$\times\,10^{14}$\,M$_\odot$.  Moreover, the two substructures are separated by 21.8\,cMpc, which is the characteristic size of a massive protocluster at $z\sim3$, and they probably merge into a more massive Coma-like galaxy cluster in the present day.

\section{Discussion} \label{sec:discussion}

\subsection{Contribution of \Hb\ and \OIII$\lambda$4959 Emission Lines} \label{sec:bias}

We take \OIII$\lambda$5007 at $z\sim 3.25$ as the emission line detected by the flux excess in the narrow band H$_2\mathrm{S}(1)$.  Given that \Hb\ and \OIII$\lambda$4959 are close to \OIII$\lambda$5007, it is possible that \Hb\ and \OIII$\lambda$4959 might contaminate our sample selection.  

We notice that the band width of H$_2\mathrm{S}(1)$ is too narrow to cover both \Hb\ and \OIII\ lines simultaneously.  \citet{Khostovan2015} pointed out that the \OIII\ line dominates the population in \Hb\ +\OIII\ luminosity function, with the fraction of \Hb\ emitters decreasing at the increasing \Hb\ +\OIII\ line luminosity. Similarly, \citet{Suzuki2016}  examined the contamination to \OIII$\lambda$5007 at $z=2.23$ and found that the contribution of \Hb\ is only $\sim$ 3$\%$ , while \OIII$\lambda$4959's contribution is up to  25\%.  Our sample \OIII\ ELGs have bright line luminosities at $\sim 3.25$. We argue that the \Hb\ contribution is negligible in our sample.  

The doublet \OIII$\lambda\lambda$4959, 5007 are separated by $\Delta\lambda=204$\,\AA\ when they redshift to $z\sim 3.25$.  The central wavelength of H$_2$S(1) redshifted to 3.25 for \OIII\ emission line.  The two lines can be both covered by the H$_2\mathrm{S}(1)$ filter ($\Delta\lambda=293$\AA) for \OIII\ emitters at $3.266<z<3.283$, compared to the redshift span of $3.225<z<3.283$ for \OIII$\lambda$5007 alone.  Therefore, our measurements of the line fluxes of \OIII$\lambda$5007 could be overestimated by 25\% owing to the contribution of \OIII$\lambda$4959 if our \OIII\ ELGs are in the redshift range of $3.266<z<3.283$.  Here a flux ratio of 3 is adopted between \OIII$\lambda$5007 and \OIII$\lambda$4959 \citep{Suzuki2016}.  We note that seven targets in our sample have spectroscopic redshifts in the range of $3.20<z<3.25$ and are spatially mixed with other sample galaxies, implying that our sample galaxies are most likely distributed at $3.225<z<3.25$ and the contribution by \OIII$\lambda$4959 should be ignorable. 
%On the other hand, the H$_2$S(1) excess selection picks the \OIII\  ELGs dominated by the \OIII$\lambda$5007 emitters \citep{Khostovan2015}.  And galaxies with strong \OIII$\lambda$4959 emission should have even stronger \OIII$\lambda$5007 emission given the line ratio.  
We thus argue that the \OIII$\lambda$5007 emission line dominates our sample galaxies. % and contamination from \Hb\ and \OIII$\lambda$4959 emitters are negligible. 
%The decreasing of spec-$z$ corroborates the fact that our sample are tend to have relatively lower redshift which means less \OIII$\lambda$4959 contribution in \OIII\ flux. Thus we suggest that \OIII\ emission line of our sample is dominated by \OIII$\lambda$5007 emission line.

\subsection{Evolution of Extreme \OIII\ ELGs over $z\sim3-4$} \label{sec:EELGs}

The vast majority of $z>3$ galaxies are star-forming, and those with strong emission lines are very common at high redshifts.  Our sample \OIII\ ELGs at $z\sim3.25$ are about 1--2\,Gyr before the cosmic star formation peak and are expected to provide clues for understanding how galaxies grow and enhance star formation.  The strong \OIII\ emission lines can be generated from the ionized regions around the hot young massive stars in a galaxy.  Galaxies with extremely strong \OIII\ emission lines at this epoch are found to have preferentially lower metallicity and higher ionization parameters powered by intense star formation activities \citep{Nakajima2014}.  The most intense \OIII\ ELGs are defined as extreme ELGs (EELGs) with a composite rest-frame EW(\OIII) of $803\pm228$\,\AA, while the less intense but still significant \OIII\ ELGs are named strong ELGs (SELGs) with a composite EW(\OIII) of $230\pm90$\,\AA\, \citep{Forrest2017}.

The \OIII\ EELGs at $z\sim3$--4, i.e., those with EW(\OIII)$_\mathrm{rest}>500$\,\AA, are widely studied in the literature \citep{Forrest2017, Cohn2018, Forrest2018, Onodera2020, Tran2020,Tang2021a,Tang2021b}. They are typically small with $M_\ast$ $\sim$ $10^{8}-10^{9}$\,M$_\odot$ and SFR $\sim$ 20--50\,M$_\odot$\,yr$^{-1}$ \citep{Maseda2014,Tran2020}.  At increasing stellar mass, EELGs tend to have higher metallicity and stronger continuum emission from evolved stellar populations.  \citet{Tran2020} demonstrated that for EELGs with higher stellar masses, their \OIII$\lambda5007$ EWs tend to decrease with relatively higher stellar continua at given star formation activities.

Our sample \OIII\ ELGs are SELGs with a median EW $\sim200$\,\AA\ in the range of $70\,\mathrm{\AA}<${EW(\OIII)}$_\mathrm{rest}<500$\,\AA.  Our results support the scenario that strong \OIII$\lambda$5007 emission reveals the early episode of intense star formation.   Our sample galaxies have stellar masses larger than $\log\,(M_\ast/$M$_\odot)\sim9$ and larger SFRs of 10--100\,M$_\odot$\,yr$^{-1}$,  denoting that our \OIII\ ELGs are more representative of the main population of SFGs than the \OIII\ EELGs with EW(\OIII)$_\mathrm{rest}>500$\,\AA.

\subsection{The \OIII\ Overdensity in ECDFS} \label{sec:ECDFS}

Protoclusters are considered as ideal laboratories to study galaxy properties in the dense environments, as well as the environmental effects on galaxy formation and evolution.  Previous studies on $z>3$ protoclusters mainly identify them with \Lya\ emitters (LAEs), Lyman break galaxies (LBGs), and submillimeter galaxies (SMGs).  Up to date, there are more than 30 protoclusters reported at $z>3$ with spectroscopically confirmed galaxies \citep[see][for a review]{Harikane2019}.  These protoclusters are likely to form ``Virgo-type'' galaxy clusters at $z=0$ with a total mass of (3--9)\,$\times\,10^{14}$M$_\odot$.

Only a few overdensity structures have been reported at $z>3$ traced with \OIII\ emitters \citep{Maschietto2008,Forrest2017}.  We show that our sample \OIII\ ELGs reside in a massive overdense structure in ECDFS.  The SE substructure spreads over an area of $7\arcmin \times 5\arcmin$ while the NW substructure covers an area of $4\farcm4 \times 4\farcm4$.  This overdensity of \OIII\ ELGs at $z\sim 3.25$ is a new structure discovered in ECDFS.  The SE and NW components have an overdensity factor about 20--60 over different comoving volumes owing to the the limits of radial comoving distance.  These two substructures are expected to be virialized at $z=0$ and probably form a massive cluster with $\sim 1.1\times10^{15}\,$M$_\odot$ for SE and $\sim 4.8\times10^{14}\,$M$_\odot$ for NW. And the two substructures probably merge into a more massive single Coma-like galaxy cluster with $\sim 2.3\times10^{15}\,$M$_\odot$.

In ECDFS, there is one overdensity traced by extreme \Hb\ + \OIII\ emitters at $z\sim 3.5$ discovered by the ZFOURGE and GOODS-ALMA surveys \citep{Forrest2017,Zhou2020}.  \citet{Forrest2017} found a redshift peak at $z=3.5$ with EELGs and SELGs in the ZFOURGE catalog.  The seventh nearest-neighbor measure is used to build the overdensities projected on the sky, revealing the densest region of extreme \OIII\ emitters in ECDFS,  with 53 member galaxies over a scale of 8.1\,cMpc.

\citet{Maschietto2008} reported 13 \OIII\ emitters around the radio galaxy MRC\,0316$-$257 at $z=3.13$. This radio-selected protocluster consists of 32 LAEs over a $7\arcmin \times 7\arcmin$ region \citep{Venemans2005}.  These 13 \OIII\ emitters form an overdensity with $\delta_{\rm gal}\sim 2.5$.  \citet{Kuiper2012} found that MRC\,0316$-$257 has a foreground structure at $z=3.10$ traced by three spectroscopically confirmed \OIII\ emitters.  They pointed out that the two structures are unlikely part of a larger protocluster based on a two-dimensional Kolmogorov--Smirnov test.  We identify two substructures in our $z=3.25$ overdense structure traced by 34 \OIII\ ELGs in ECDFS.  We lack spectroscopic redshifts to see whether the NW substructure is located at the same redshift as the SE substructure.  From the spatial distribution of our 34 \OIII\ ELGs, the two substructures likely belong to the same large-scale structure.

A recent work from the VANDELS survey presented several overdensities traced by Lyman-$\alpha$ emitters at $2<z<4$ in CDFS and UDS \citep{Guaita2020}.  We take their overdensities near $z=3.25$ for a comparison.  As shown in Figure~\ref{f:SD.pdf}, the central locations of three overdensities at $z$=3.17, 3.23, and 3.29 are overplotted in ECDFS.  However, the spatial locations of the VANDELS \Lya\ overdensities are not exactly coincident with our SE and NW components.  The spatial offsets between them could be explained by the systematic offsets between different populations from the density tracers, say, the \OIII\ and \Lya\ emitters.  \OIII\ emitters are more massive and metal rich, while the \Lya\ emitters are mostly low-mass and metal poor.  The typical present-day mass of the VANDELS overdensities is about $0.3\,\times\,10^{13}$\,M$_\odot$, which is about a factor of two and three lower than our NW and SE components, respectively.

%The spatial distribution of \Ha\ emitters in ECDFS at $z=2.24$ \citepalias{An2014} is also compared to the one of \OIII\ emitters at $z=3.25$.  We find that the central region of SE component has no \Ha\ emitters distributed in the foreground, showing no connection between \Ha\ emitters and the \OIII\ overdensity.  Due to the lack of deep $U$ and $I$ data or spectroscopic observations, the \Lya\ emission information is unknown in ECDFS.  Among the 13 \OIII\ emitters in \citet{Maschietto2008}, 5 of them are also confirmed with \Lya\ emission while 3 are newly detected.  The connection of \OIII\ emitters and \Lya\ emitters are not fully understood in high-$z$, a \Lya\ emission line detection is needed for ECDFS to have a better study on this problem.    

In addition, some of our sample \OIII\ ELGs exhibit significant dust attenuation and high SFR, compared to normal SFGs at the same redshifts.  This hints that star formation and metal enrichment in this overdensity are enhanced.  No detection of extreme \OIII\ ELGs (EW(\OIII)$_\mathrm{rest}>500$\,\AA) in this overdensity also supported the acceleration of galaxy evolution in the overdense environment, in which low-mass and low-metallicity starburst galaxies are deficient.  However, protoclusters at similar redshifts have been found to have quiescent galaxies largely concentrated in the overdense region, probably due to the environmental quenching \citep{Shi2021,McConachie2022}.  These hint that the evolutionary states of protoclusters largely decide the environmental impacts on the member galaxies at $z>3$.

\section{Summary and Conclusions} \label{sec:conclusion}

Using the deep narrowband H$_2\mathrm{S}(1)$ and broadband $K_\mathrm{s}$ imaging of ECDFS, we identify a sample of 34 \OIII\ ELGs at $z\sim3.25$ and carry out an analysis of their physical properties.  Using preexisting multiwavelength data, we construct SEDs from $U$ to $K_\mathrm{s}$ and perform SED fitting with CIGALE to obtain rest-frame $UVJ$ colors,  stellar mass, SFR, dust attenuation, \OIII\ EW and \OIII\ luminosities.  The sample \OIII\ ELGs map an overdense structure that possible be the progenitor of a Coma-like massive galaxy cluster at $z\sim0$.  Our main results are summarized as follows: 

\begin{enumerate}

\item The vast majority of our sample are SFGs with strong \OIII\ emission lines. Compared with the extreme \OIII\ ELGs of EW(\OIII)$_\mathrm{rest}>500$\,\AA, our sample galaxies have EW(\OIII)$_\mathrm{rest}\sim$ 70--500\,\AA\ and are more massive with  $M_{\ast}\sim$\,10$^{9.0}-10^{10.6}$\,M$_\odot$, and higher SFRs of $\sim$ 10--210 M$_\odot$\,yr$^{-1}$ compared to typical SFGs at the same masses.  Only one target is identified as an  AGN detected in the \textit{Chandra} 7\,Ms X-ray observations.  According to the $UVJ$ color-color diagram, the majority of our \OIII\ ELGs are located at the blue end of the star-forming regime.  Our sample \OIII\ ELGs exhibit significant dust attenuation and high SFR compared to normal star-forming galaxies.  Our results show that the NIR selection is able to pick a significant fraction of \OIII\ ELGs with high dust attenuation ($A_V>1$\,mag), and our sample is more representative of the main population of SFGs at $z\sim3$--4 in comparison with the extreme \OIII\ ELGs with EW(\OIII)$_\mathrm{rest}>500$\,\AA.
\item With \textit{HST}/ACS and WFC3 observations we show that $z=3.25$ \OIII\ ELGs have a variety of morphologies in the rest-frame UV and optical.  We find that 38\% (13/34) of our \OIII\ ELGs appear to have a diffuse/clumpy/tidal shape and about 35\% (12/34) are very compact with $R_\mathrm{e}<0\farcs3$.  Three \OIII\ ELGs are identified as mergers, and four are considered to be galaxy pairs with two galactic nuclei with similar color and size within a separation distance of $2\arcsec$.  Two \OIII\ ELGs are too faint in the rest-frame UV to be recognized in morphology, likely due to heavy dust attenuation.  And the large fraction of compact sources in our sample also are tend to be located above the star formation main sequence of $z\sim3$ SFGs.
\item We find that our \OIII\ ELGs trace an overdense structure at $z=3.25$.  This structure is composed of two substructures of scales of $5\arcmin\times7\arcmin$ and $4\farcm4\times4\farcm4$, separated by 21.8\,cMpc.  We take the half of narrowband filter redshift span as the upper limit and the minor axis of the three different overdensity rectangular areas as the lower limit for the line-of-sight comoving distance to estimate the number density per comoving volume.  Our estimate suggests that this structure has an overdensity factor $\delta_\mathrm{gal} \sim$ 9--12 over a comoving volume of 25$^3$--23$^3$\,cMpc$^3$.  The SE and NW substructures are denser with $\delta_\mathrm{gal}$ in the range of 22--60 over a  volume of 15$^3$--11$^3$\,cMpc$^3$, and 18--57 over a volume of 12$^3$--8$^3$\,cMpc$^3$, respectively.  We estimate their present-day mass to be $\sim1.1\times10^{15}\,$M$_\odot$ for SE and $\sim4.8\times10^{14}\,$M$_\odot$ for NW, and these two substructures are likely to merge into a Coma-like massive cluster with $\sim 2.3\times10^{15}\,$M$_\odot$ at the present day. % By comparison of overdensity discovered in CDFS at $z\sim3.5$ \citep{Forrest2017}, we suggest that these two confirmed overdensities indicate a large cosmic web with strong \OIII\ emission in CDFS at $z\sim3.25$--3.5.
\item   We stress that none of our sample \OIII\ ELGs exhibit EW(\OIII)$_\mathrm{rest}>500$\,\AA. We argue that the lack of \OIII\ EELGs is largely due to our sample of \OIII\ ELGs residing in the overdense environment, in which star formation and chemical enrichment in galaxies are enhanced and  low-mass and low-metallicity starburst galaxies are deficient.  

\end{enumerate}

\begin{acknowledgments}

%We appreciate those who provide valuable comments and suggestions that improved our manuscript.
This work is supported the National Science Foundation of China (12073078 and 12173088); the Major Science and Technology Project of Qinghai Province (2019-ZJ-A10); the science research grants from the China Manned Space Project with Nos. CMS-CSST-2021-A02, CMS-CSST-2021-A04 and CMS-CSST-2021-A07, 
and the Chinese Academy of Sciences (CAS) through a China-Chile Joint Research Fund (CCJRF No.1809) administered by the CAS South America Centre for Astronomy (CASSACA). 

This research adopts data obtained through the Telescope Access Program (TAP), 
which is funded by the National Astronomical Observatories and the Special Fund for Astronomy from the Ministry of Finance. 

\end{acknowledgments}

\end{document}